\documentclass{IEEE-Con-Sys-mag}
\pdfoutput=1

\title{Statistical Learning Theory for Control\stitle{A Finite Sample Perspective}}
\author{Anastasios Tsiamis*, Ingvar Ziemann*, Nikolai Matni, and George J. Pappas}
\affil{A. Tsiamis (\href{mailto:atsiamis@control.ee.ethz.ch}{atsiamis@control.ee.ethz.ch}) is with the Dept. of Information Technology and Electrical Engineering, ETH Zürich, Zürich, Switzerland.\\
I. Ziemann (\href{mailto:ziemann@kth.se}{ziemann@kth.se}) is with the Division of Decision and Control Systems, KTH Royal Institute of Technology, Stockholm, Sweden.\\
N. Matni (\href{mailto:nmatni@seas.upenn.edu}{nmatni@seas.upenn.edu}) and G. J. Pappas (\href{mailto:pappasg@seas.upenn.edu}{pappasg@seas.upenn.edu}) are with the Dept. of Electrical and Systems Engineering, University of Pennsylvania, Philadelphia, USA.\\
*Both authors contributed equally.}
\usepackage{amsthm}
\usepackage{amssymb}
\usepackage{amsmath}
\usepackage{mathrsfs}
\usepackage{hyperref}
\usepackage{makecell}

\usepackage[most]{tcolorbox}
\usepackage{graphicx}
\hypersetup{
    unicode=false,          
    pdftoolbar=true,        
    pdfmenubar=true,        
    pdffitwindow=false,     
    pdfstartview={FitH},    
    pdfnewwindow=true,      
    colorlinks=true,       
    linkcolor=blue,          
    citecolor=blue,        
    filecolor=magenta,      
    urlcolor=red,           
    breaklinks=true
}
\usepackage{cleveref}
\usepackage{xcolor}
\usepackage[sort, numbers, compress]{natbib}
\usepackage{here}
\usepackage{caption}
\usepackage{subcaption}

\newcommand{\dx}{d_{\mathsf{x}}}
\newcommand{\du}{d_{\mathsf{u}}}
\newcommand{\dy}{d_{\mathsf{y}}}
\newcommand{\dz}{d_{\mathsf{z}}}
\newcommand{\dhh}{d_{\eta}}
\newcommand{\dd}{d}
\newcommand{\dsparse}{d_{\mathsf{s}}}
\newcommand{\snr}{\mathrm{snr}}
\newcommand{\poly}{\mathrm{poly}}
\newcommand{\Ntraj}{N_\mathrm{traj}}
\newcommand{\Ntot}{N_\mathrm{tot}}
\newcommand{\As}{A_{\star}}
\newcommand{\Bs}{B_{\star}}
\newcommand{\Cs}{C_{\star}}
\newcommand{\thetas}{\theta_{\star}}
\newcommand{\Lk}{L_\star}
\newcommand{\Hs}{\tilde{B}_{\star}}
\newcommand{\Hk}{\mathcal{H}}
\newcommand{\obs}{\mathcal{O}}
\newcommand{\contr}{\mathcal{C}}
\newcommand{\Acls}{A_{\mathrm{cl},\star}}
\newcommand{\Ss}{S_\star}
\newcommand{\Contr}{\mathcal{K}}
\newcommand{\Regret}{\mathcal{R}}
\newcommand{\E}{\mathbf{E}}
\newcommand{\F}{\mathcal{F}}
\newcommand{\Real}{\mathbb{R}}
\newcommand{\Sigmaw}{\Sigma_w}
\newcommand{\Sigmav}{\Sigma_v}
\newcommand{\Sigmae}{\Sigma_e}

\newcommand{\set}[1]{\ensuremath{\left\{ #1\right\}}}

\newcommand{\e}{\epsilon}

\newcommand{\Prob}{\mathbf{P}}
\newcommand{\V}{\mathbf{V}}

\newcommand{\CC}{\ensuremath{\mathscr{C}}}

\DeclareMathOperator{\argmin}{argmin}

\DeclareMathOperator{\VEC}{\mathsf{vec}}

\DeclareMathOperator{\I}{\mathtt{I}}
\DeclareMathOperator{\J}{\mathtt{J}}

\newcommand{\norm}[1]{\lVert #1 \rVert}

\newcommand{\opnorm}[1]{\| #1 \|_{\mathsf{op}}}

\newcommand{\clint}[1]{\ensuremath{\left[ #1\right]}}
\renewcommand{\set}[1]{\ensuremath{\left\{ #1\right\}}}
\newcommand{\matr}[1]{\ensuremath{\clint{\begin{array} #1 \end{array}}}}

\newcommand{\rev}[1]{{\color{black} #1}}
\newtheorem{problem}{Open Problem}
\newenvironment{open_prob}
  { \begin{tcolorbox}[
 colframe=yellow!70!white,
 colback=yellow!17!white,
 arc=8pt,
 breakable,
 left=1pt,right=1pt,top=1pt,bottom=1pt,
 boxrule=0.3pt,
 ]
\begin{problem}}
  {\end{problem}
 \end{tcolorbox}}

\begin{document}

\maketitle

\chapterinitial{L}earning algorithms have become an integral component to modern engineering solutions. Examples range from self-driving cars, recommender systems, finance and even critical infrastructure, many of which are typically under the purview of control theory. While these algorithms have already shown tremendous promise in certain applications \cite{silver2016mastering} there are considerable challenges, in particular with respect to guaranteeing safety and gauging fundamental limits of operation. Thus, as we integrate tools from machine learning into our systems, we also require an integrated theoretical understanding of how they operate in the presence of dynamic and system-theoretic phenomena. 

Over the past few years, intense efforts toward this goal---an integrated theoretical understanding of learning, dynamics and control---have been made. While much work remains to be done, a relatively clear and complete picture has begun to emerge for (fully observed) linear dynamical systems. These systems already allow for reasoning about concrete failure modes, thus helping to indicate a path forward. Moreover, while simple at a glance, these systems can be challenging to analyze. Recently, a host of methods from learning theory and high-dimensional statistics, not typically in the control-theoretic toolbox, have been introduced to our community.

\paragraph{Outline}
This tutorial survey serves as an introduction to these results for learning in the context of  unknown linear dynamical systems. We review the current state of the art and emphasize which tools are needed to arrive at these results. Our focus is on characterizing the sample efficiency and fundamental limits of learning algorithms. Along the way, we also delineate a number of open problems. More concretely, this paper is structured as follows: We begin by revisiting recent advances in the \nameref{sec:finsampsysid}. Next, we discuss how these finite sample bounds can be used downstream to give guaranteed performance for learning-based \nameref{sec:offlinecontrol}. The last technical section discusses the more challenging \nameref{sec:regmin} setting. Finally, in light of the material discussed, we outline a number of \nameref{sec:futdir}.

\begin{summary}
\summaryinitial{T}his tutorial survey provides an overview of recent  advances in statistical learning theory relevant to control and system identification featuring non-asymptotics. While there has been substantial progress across all areas of control, the theory is most well-developed when it comes to linear system identification and learning for the linear quadratic regulator, which are the focus of this manuscript. From a theoretical perspective, much of the labor underlying these advances has been in adapting tools from modern high-dimensional statistics and learning theory. While highly relevant to control theorists interested in integrating tools from machine learning, the foundational material has not always been easily accessible. To remedy this, we provide a self-contained presentation of the relevant material, outlining all the key ideas and provide an overview of the technical machinery that underpin recent results. We also present a number of open problems and future directions.
\end{summary}

\section{finite sample analysis of system identification}
\label{sec:finsampsysid}
In linear system identification, the goal is to recover the model of an \emph{unknown} system of the form~\eqref{ID_eq:system} below:
 \begin{equation}\label{ID_eq:system}
\begin{aligned}
    x_{t+1}&=\As x_{t}+\Bs u_{t}+w_t\\
    y_{t}&=\Cs x_t+v_t\,,
\end{aligned}
\end{equation} 
where $x_t\in\Real^{\dx}$ represents the state, $y_t\in\Real^{\dy}$ represents the observations, $u_t\in\Real^{\du}$ is the control signal, and $w_t\in\Real^{\dx}$, $v_t\in\Real^{\dy}$ are the process and measurement noises respectively.

The question we answer in this section is ``\emph{how many samples are needed to guarantee that system identification error is small}"? We will make this question more formal by introducing the notion of \emph{sample complexity}.  Prior to doing so, we establish the statistical learning framing of the problem.

While many of the results presented in the following subsections can be extended to more general noise models, we keep the exposition simple by focusing on Gaussian noise models.  In particular, we assume that both the process noise $w_t$ and measurement noise $v_t$ are i.i.d. zero mean Gaussians with covariance matrices $\Sigmaw$ and $\Sigmav$ respectively, and that these process are all mutually independent of each other.  Similarly, we let the initial state $x_0$ be a zero mean Gaussian, with covariance $\Gamma_0$, and independent of the process and measurement noise. We denote the covariance of the state $x_t$ at time $t$ by $\Gamma_{t}\triangleq \E x_t x^{\top}_t$.

Here and in the sequel, the state parameters $(\As,\Bs,\Cs)\in\Real^{\dx\times(\dx+\du+\dy)}$ are unknown. The goal of the system identification problem is to recover the a priori unknown model of system~\eqref{ID_eq:system} from finite input-output samples $\{(y_i,u_i)\}_{i=1}^{\Ntot}$, where $\Ntot$ is the total number of samples. As such, this is an \emph{offline learning} problem.  The data can come from a single trajectory of length $T$, i.e., $N_\mathrm{tot}=T$, or come from $\Ntraj$ multiple \rev{independent} trajectories with horizon $T$, i.e., $\Ntot=T\Ntraj.$ While the learning task is to recover the state-space parameters $\thetas\triangleq (\As,\Bs,\Cs,\Sigmaw,\Sigmav)$ of~\eqref{ID_eq:system} using this data, the state-space representation of system~\eqref{ID_eq:system} is in general not unique.  As such, we instead seek to recover one such representation or a function $f(\thetas)$ of the underlying true parameters $\thetas$. To streamline exposition, we focus on the single trajectory case $N_{\mathrm{tot}}=T$. A more refined analysis can be used when samples are drawn from multiple trajectories to yield similar conclusions \cite{tu2022learning} but under weaker stability-type assumptions.

Let the identification algorithm $\mathcal{A}$ be a (measurable) function that takes as an input the horizon $T$ and the data $\{(y_0,u_0),(y_1,u_1),\dots,(y_T,u_T)\}$, and returns an estimate $\widehat{f}_T$ of the desired system quantity $f(\thetas)$. 
In some settings, the algorithm $\mathcal{A}$ may also encompass an exploration policy, i.e., the choice of control inputs $u_t$ used during the data-collection phase.  \rev{
The goal of the exploration policy is to excite the system in a way that maximizes the ``richness" of the data, that is, how much information the data carry about the underlying system.}
Formally, we define an exploration policy $\pi$ to be a sequence of (measurable) functions $\pi=\set{\pi_t}_{t=0}^{\infty}$, where every function $\pi_t$ maps previous output-input values $y_0,\dots,y_t,u_0,\dots,u_{t-1}$ and potentially an auxiliary randomization signal to the new input $u_t$. This definition encompasses both closed and open-loop policies---in the latter case, the exploration policy is only a function of the auxiliary randomness.

We can now define the notion of sample complexity.  
Let $\Prob_{\theta,\pi}$ denote the probability distribution of the input-output data for the system~\eqref{ID_eq:system} defined by parameters $\thetas$ evolving under the exploration policy $\pi$. 
 \begin{tcolorbox}[
 colframe=red!70!white,
 colback=red!9!white,
 arc=8pt,
 breakable,
 left=1pt,right=1pt,top=1pt,bottom=1pt,
 boxrule=0.3pt,
 ]
\textbf{Sample Complexity.} Fix a class $\CC$ of systems of the form~\eqref{ID_eq:system} and a norm $\norm{\cdot}$. Let $f(\thetas)$ be the system quantity to be identified. Fix an identification algorithm $\mathcal{A}$ with an exploration policy $\pi$. Pick an accuracy parameter $\varepsilon$ and a failure probability $\delta\in(0,1)$.  Let $\widehat{f}_T$ be the system identification output under the algorithm $\mathcal{A}$. Then the \textbf{sample complexity} $N_c$ of learning $f$ given the class $\CC$, the algorithm $\mathcal{A}$, and the policy $\pi$ is the minimum $N_c=N_c(\varepsilon,\delta,\CC,\mathcal{A},\pi)$ such that:
 \begin{equation}\label{ID_eq:sample_complexity_algorithm}
\begin{aligned}
		&\sup_{\thetas\in\CC}\Prob_{\thetas,\pi}(\norm{f(\thetas)-\widehat{f}_T}\ge \varepsilon)\le \delta\\
		&\text{ if }T\ge N_c(\varepsilon,\delta,\CC,\mathcal{A},\pi).
	\end{aligned}
 \end{equation}
 We say that a class of systems $\CC$ is \textbf{learnable} if there exist an algorithm $\mathcal{A}$ and a policy $\pi$ such that for any $\varepsilon>0$, $\delta \in (0,1)$ the sample complexity $N_c$ is finite.
 \end{tcolorbox}

\begin{sidebar}{What do finite-sample methods bring?}
\section[what do finite-sample methods bring?]{}\phantomsection
   \label{sidebar-CLT}
\setcounter{sequation}{0}
\renewcommand{\thesequation}{S\arabic{sequation}}
\setcounter{stable}{0}
\renewcommand{\thestable}{S\arabic{stable}}
\setcounter{sfigure}{0}
\renewcommand{\thesfigure}{S\arabic{sfigure}}
\rev{\sdbarinitial{C}onsider an \emph{unknown} scalar system
\begin{sequation}\label{eq:simple_example}
x_{t+1}=a_{\star}x_t+w_t,
\end{sequation}
where $|a_\star|<1$, $w_t$ is i.i.d. and mean-zero Gaussian with variance $1$. Assume that our goal is to recover the unknown scalar $a_\star$ from single trajectory data $(x_0,\dots,x_{T})$. One of the simplest algorithms is to minimize the squared prediction errors
\[
\hat{a}_T=\arg\min_a \sum_{t=1}^{T}(x_{t}-ax_{t-1})^2.
\]
Given the stochastic nature of the data, the least-square estimate $\hat{a}_T$ will fluctuate around the ``true" value $a_\star$. 

Both asymptotic and non-asymptotic methods aim to characterize the statistical variability of the error $\hat{a}_T-a_\star$.
One of the most powerful \textbf{asymptotic} tools is establishing asymptotic normality, i.e. a time-series version of the Central Limit Theorem (CLT). For this particular scalar system, Mann and Wald~\cite{mann1943statistical} proved that as the number of samples approaches infinity $T\rightarrow \infty$, the estimation error is asymptotically normal
\[
\sqrt{T}(\hat{a}_T-a_\star)\Rightarrow \mathcal{N}(0,1-a^2_\star),
\]
where $\Rightarrow$ denotes convergence in distribution and $\mathcal{N}(\mu,\sigma^2)$ denotes the normal distribution with mean $\mu$ and variance $\sigma^2$. This result can give us the sharpest bound in the asymptotic regime. However, it requires an infinite number of samples and can only be used as a heuristic under finite samples. 
Some questions remained unanswered. For example, what is the distribution of the error under finite samples? What is the transient behavior? 

We can partially answer these questions by applying the \textbf{non-asymptotic} tools reviewed in this survey. In particular, by following the arguments in~\nameref{sec:samplecomplexityub}, we can establish a finite-sample tail bound of the form
\[
\Prob(|\hat{a}_T-a_\star|\ge \varepsilon)\le \delta,
\]
for a large enough sample size
\[
T\ge \max\left\{T_{\mathrm{burn-in}},c \frac{1-a^2_\star}{\varepsilon^2}\log\frac{ 1}{\delta}\right\},
\]
where $\varepsilon$ controls the accuracy of identification and $\delta$ controls the confidence. The constant $c$ is a so-called ``universal" constant, i.e. it just takes a numerical value and is independent of system parameters, confidence, and accuracy. The burn-in time $T_{\mathrm{burn-in}}$ captures the complexity of transient phenomena, e.g. the minimum time until we achieve persistency of excitation (excitation of all modes of the system). It typically depends on the desired confidence but not on the accuracy $\varepsilon$. For the simple scalar system~\eqref{eq:simple_example}, we can take $T_{\mathrm{burn-in}}=c' \log 1/\delta$, where $c'$ is another universal constant.

While we did not fully characterize the finite-sample distribution of the estimation error, we managed to characterize the tail probabilities. For example, we have a $\log1/\delta$ term in the required number of samples, which is sharp. This was not possible before by applying only asymptotic tools--see~\cite[Ch 2.1]{vershynin2018high} for a more technical explanation.  We can even achieve a finite-sample bound for $a_\star=1$ (not presented in this sidebar), when the system does not converge to a steady-state distribution. Similar properties hold in the case of general vector-valued systems of high dimensions $\dx>1$. In fact, we can even allow the state dimension $\dx$ to increase with the number of samples $T$, which is not covered by CLT.

A downside of finite-sample bounds is that we lose sharpness in the asymptotic regime. In particular, the universal constants $c,c'$ (see~\cite{simchowitz2018learning} for exact expressions) are typically large numerical values, much larger than the ones that we would obtain from a heuristic application of CLT. Nonetheless, non-asymptotic bounds can provide a detailed qualitative characterization of learning complexity. }
\end{sidebar}

In the case of multiple trajectories, we can replace $T$ with $\Ntot$ in the above definition.
We can also define algorithm-independent and/or policy-independent sample complexity, by considering the minimum $N_c$ over all possible algorithms/policies.  By choosing $\CC$ to be a neighborhood around some system $\thetas$, we can also define local, instance-specific sample complexities, see for example~\cite{jedra2019sample}. Note that for the sample complexity to be non-trivial, the algorithm should perform well across all possible $\thetas \in\CC$, which is what the supremum over $\CC$ achieves in~\eqref{ID_eq:sample_complexity_algorithm}.  Otherwise, we can construct trivial algorithms that overfit to a specific system and fail to identify any other system in the class.  

\rev{Let us also point out that one often encounters ranges of $T$ and $\delta$ for which the sample complexity dependency on $\varepsilon$ behaves poorly. Typically, this is due to transient phenomena. For instance, in a $d$-dimensional linear regression problem, the design matrix can be near singular if we have too few measurements (e.g. if fewer than $d$ independent measurements are available). Informally, for a fixed $\delta$, one typically refers to the smallest sample size $T$ such that there exists a finite (or meaningful) sample complexity at accuracy $\varepsilon$ as the \emph{burn-in time}. The burn-in for linear system identification is given in \eqref{ID_eq:Burn_In}. }

\subsection{From Asymptotics to Finite Sample Guarantees}
Before we proceed let us take a step back and briefly discuss the historical development of system identification from a mathematical methods perspective. Clearly, the statistical analysis of system identification algorithms has a long history~\cite{Ljung1999system}. Until recently, this line of work has emphasized providing guarantees for  system identification algorithms in the \emph{asymptotic regime}~\cite{lai1983asymptotic,ljung1992asymptotic,deistler1995consistency,bauer1999consistency,chiuso2004asymptotic},  in which the number of collected samples tends to infinity. The main focus of asymptotic analysis has been to establish consistency, i.e., the convergence of the estimated system parameters to the ground truth (as modelled).  Typically this is achieved if certain \emph{persistency of excitation} conditions hold~\cite{bai1985persistency}. Asymptotic tools can also go beyond consistency and provide convergence rates.
Standard tools for characterizing such rates are the Law of Iterated Logarithm (LIL) and the Central Limit Theorem (CLT)---see~\cite{hannan2012statistical} for a detailed exposition of both techniques. Nevertheless, even the more advanced techniques, i.e. the LIL and the CLT, only hold as the number of samples tend to infinity.

\paragraph{Toward a finite sample analysis}
Early work on the non-asymptotic analysis of system identification appeared in the 90s~\cite{dahleh1993sample,poolla1994time,guo1995performance,goldenshluger1998nonparametric,weyer1999finite} and 00s~\cite{campi2002finite,vidyasagar2008learning}. The setting of~\cite{dahleh1993sample,poolla1994time} focuses on worst-case noise, which is different from the statistical setting considered in this survey. \rev{In~\cite{guo1995performance} approximate expressions for the finite-time identification-error variance are given. We cannot derive sample complexity guarantees directly from~\cite{guo1995performance}; the expressions therein are not directly computable in our setting, and  they do not characterize the finite-sample distribution of the identification error and how it depends on the number of samples. 
}
The statistical learning setting was first studied in~\cite{weyer1999finite,campi2002finite,vidyasagar2008learning}, where the guarantees are typically given for the prediction error of the learned model.  Moreover, the guarantees rely heavily on having a mixing, i.e., a stable, process. As we will soon see, in many settings mixing is not required, and in fact faster mixing systems can be harder to learn---at least when it comes to parameter recovery~\cite{simchowitz2018learning}.
Following the papers by~\citet{abbasi2011regret} and~\citet{dean2020_on_the_sample}, there has been a resurgence of interest in using finite data tools for system identification and controls. This is partially motivated by  recent advances in high-dimensional probability~\cite{vershynin2018high} and statistics~\cite{wainwright2019high}, which provide us with new, powerful, tools and allow us to bypass asymptotic reasoning. 

\rev{
\paragraph{Why do we need finite sample guarantees?}
 In principle, our view is that both asymptotic and non-asymptotic methods are useful for both control and learning theorists to have in their toolbox.   
On the one hand, a careful asymptotic analysis can provide sharp bounds and give a clear picture of some key quantities involved in the problem at hand. However, in reality, all data is finite, and asymptotic bounds are heuristics, albeit often sharp if the sample size is large enough. On the other hand, non-asymptotic analysis is often more appropriate to carefully delineate notions such as transient phenomena (e.g. burn-in times) and failure probabilities--see \nameref{sidebar-CLT}. We gain a more detailed qualitative characterization of learning difficulty, often at the expense of sharpness in the asymptotic regime. For instance, the question "how many samples do we need to stabilize an unknown linear system with a certainty equivalent LQR controller?" is necessarily answered using finite sample methods. Being able to combine these sometimes distinct styles of analysis gives us a richer understanding of the dynamic phenomena under consideration.

Many datasets are high-dimensional with the number of explanatory variables not necessarily being small in proportion to the number of samples collected, e.g. the state dimension $\dx$ might be of the same order as $T$. In this case, asymptotic bounds with fixed dimension $\dx$ are not always meaningful, while finite-sample guarantees still hold. Examples from systems theory for when this may be relevant include large networked system and auto-regressions of unknown order. An insightful discussion on this matter from a statistics perspective is held by \citet[Chapter 1]{wainwright2019high}.

From the perspective of a \textbf{control theorist}, obtaining sample complexity bounds as a function of system theoretic parameters, e.g. system dimension, controllability gramian, stability radius, etc, could be very useful. Finite-sample bounds can be qualitatively informative about learning difficulty and what can go wrong with it. We can answer questions like ``which systems are hard to learn?", ``how does the controllability structure affect learnability?", ``which algorithms are optimal?". Naturally, some of these questions can also be answered using asymptotic tools. Nonetheless, we believe that a finite-sample approach offers a new perspective, giving us tools to even pose new questions--see, for instance, the open problems later. 

Learning control systems is also interesting from the perspective of a \textbf{machine learning theorist}. While the setting of learning under independent or weakly-dependent (mixing) data has been studied extensively, new challenges arise in control systems, where the data are not only dependent but also affected by control inputs. Some questions that are of interest are ``when is learning under dependent data as easy as learning under independent data?", "is mixing  required?", ``what is the tradeoff between exploration and exploitation?".

Lastly, a goal of this survey is to \textbf{establish a common language} between control theorists, learning theorists and statisticians. Machine learning theory has in principle been non-asymptotic from the outset and modern statistics has very much moved in this direction. Meanwhile, the classical literature of system identification and adaptive control relies, more often than not, on asymptotic tools. A common language facilitates an exchange of ideas that is likely to benefit all three fields. Besides, Machine Learning, Statistics and Control Theory share common research agendas and often seek to tackle the same problems.

\paragraph{Asymptotic Notation}
In this paper, we sometimes use the asymptotic notation $O,\Theta,\Omega$ to simplify the presentation. This does not imply that our statements are asymptotic. For example, the statement $f(T)=O(g(T))$ ( $f(T)=\Omega(g(T))$) can be  replaced by statements of the form ``there exists universal positive constant $c>0$ such that $f(T) \leq c g(T)$ ($f(T) \geq c g(T)$), for $T \geq T_{\mathrm{burn-in}}$", where a universal constant just takes a numerical value and is independent of system and algorithmic parameters. Exact finite-time expressions for $g(T),c,T_{\mathrm{burn-in}}$ are given either here, e.g. see~\eqref{ID_eq:confidence_ellipsoids}, or in the respective papers. 
The statement $f(T)=\Theta(g(T))$ is equivalent to $f(T)=O(g(T)),\,f(T)=\Omega(g(T))$ holding simultaneously. Lastly, the $\tilde{O}$ notation ignores poly-logarithmic terms, e.g. $f(T)=\tilde{O}(g(T))$ is equivalent to $f(T)=O(g(T) \mathrm{poly}(\log T))$, where $\mathrm{poly}$ denotes some arbitrary polynomial function of fixed degree.
}

\subsection{Fully Observed Systems}\label{sec:fully_observed}
Let us now return to the technical task at hand: to provide a finite sample analysis system identification. Recall that we focus on the single trajectory case $\Ntot=T$. We start by analyzing the simplest system identification, namely the case of fully observed systems with $\Cs=I$ and $\Sigmav=0$, yielding direct state measurements $y_t=x_t$, $t\le T$. We will only focus on identification of $\As,\Bs$, but the same techniques could be applied for the estimation of the covariance $\Sigmaw$. For this reason, abusing the notation introduced above, we will denote $\thetas=(\As,\Bs)$, $f(\theta)=\theta$. Given the data $\{(x_0,u_0),\dots,(x_T,u_T)\}$, a natural way to obtain an estimate of the system matrices is to employ the least squares algorithm
\begin{equation}\label{ID_eq:ls_algorithm}
    \widehat \theta_T    \triangleq (\widehat A_T, \widehat B_T) \in \argmin_{A,B} \sum_{t=0}^{T-1} \|x_{t+1}-Ax_t-Bu_t\|_2^2.
\end{equation}
After some algebraic manipulations, we can verify that
{\medmuskip=0mu\thickmuskip=1mu\thinmuskip=5mu
\begin{align}\arraycolsep=1.4pt
\label{ID_eq:LSEinAC}
    \widehat \theta_T  -\thetas = \left(\sum_{t=0}^{T-1} w_{t} \begin{bmatrix} x_t^\top & u_t^\top \end{bmatrix} \right)\left( \sum_{t=0}^{T-1} \begin{bmatrix} x_t \\ u_t \end{bmatrix}\begin{bmatrix} x_t^\top & u_t^\top \end{bmatrix} \right)^{-1}
\end{align}
provided the matrix inverse on the right hand side of equation (\ref{ID_eq:LSEinAC}) exists.}

We characterize the sample complexity of the least squares estimator \eqref{ID_eq:LSEinAC} by establishing bounds on the operator norm $\opnorm{\widehat \theta_T-\thetas}$. It is possible to provide similar guarantees for the Frobenius norm, but dimensional factors differ slightly. The techniques presented below can be applied to open-loop non-explosive systems, when all the eigenvalues of matrix $\As$ are inside or on the unit circle, i.e. $\rho(\As)\le 1$, where $\rho(\As)$ denotes the spectral radius. We also assume that the open-loop inputs are i.i.d. zero-mean Gaussians with $\E u_t u^{\top}_t=\sigma^2_u I$, for some $\sigma_u>0$. We will discuss generalizations later on.
To simplify the exposition, we will also assume that the noise is full rank, i.e., $\Sigmaw\succ 0$. 
This implies that the noise directly excites all system states directly, making persistency of excitation easier to establish. We can also obtain persistency of excitation for indirectly excited systems, as long as the controllability structure of the system is well-defined~\cite{tsiamis2021linear}. Finally, we assume that the system starts from the fixed initial condition $x_0=0$, and hence the initial state-covariance is $\Gamma_0=0$.

The following terms will be useful in the analysis of the least squares algorithm
\begin{equation}\label{ID_eq:S_and_V}
S_T\triangleq \sum^{T-1}_{t=0} w_t  \begin{bmatrix} x_t^\top & u_t^\top \end{bmatrix},\,V_T\triangleq \sum_{t=0}^{T-1} \begin{bmatrix} x_t \\ u_t \end{bmatrix}\begin{bmatrix} x_t^\top & u_t^\top \end{bmatrix}.
\end{equation}
Using the above notation, we can break the least-squares error into two separate terms
\[
\opnorm{\widehat{\theta}_T-\thetas}
\le\underbrace{\opnorm{S_TV^{-1/2}_T}}_{\text{Self-normalized term}}\underbrace{\opnorm{V^{-1/2}_T}}_{\text{PE term}},
\]
where $V^{-1/2}_T$ denotes a symmetric positive definite matrix such that $V^{-1/2}_TV^{-1/2}_T=V^{-1}_T$.
To obtain sample complexity bounds for the least square algorithm, we need to analyze both terms. The self-normalized term captures the contribution of the noise to the least squares error. The PE term captures Persistency of Excitation (PE), i.e., the richness of the data. The richer the data, the larger the magnitude of the eigenvalues of the Gram matrix $V_N$, leading to a smaller identification error.
\subsubsection{Persistency of excitation}
If the collected trajectory data is rich enough, i.e., if all modes of the system are excited, then the gram matrix $V_T$ defined in \eqref{ID_eq:S_and_V} is both invertible and well-conditioned. In particular, if $\lambda_{\min}(V_T)$ grows unbounded with $T$, we say that Persistency of Excitation (PE) holds. Moreover, the smallest eigenvalue of $V_T$ captures the direction of the system which is the most difficult to excite.

Recall that $\Gamma_t=\E x_tx^\top_t$ is the covariance of the state. Under i.i.d. white inputs, we can compute
\begin{equation}
\Gamma_t=\sum_{k=0}^{t-1}A^k(\sigma^2_uBB^\top+\Sigmaw)(A^{\top})^k ,\,\Gamma_0=0. \label{eq:gramdef}
\end{equation}
Since the state is driven by both exogenous inputs and noise, both factors appear in the state covariance. By the definition of the Gram matrix $V_T$, we have
\[
\E V_T=\begin{bmatrix}
\sum^{T}_{t=0}\Gamma_t&0\\0&\sigma^2_u TI
\end{bmatrix}.
\]
Note that $\Gamma_t$ is increasing in the positive semi-definite cone, since $\Gamma_0=0$.
It is easy to show that the expected Gram matrix $\E V_T$ is invertible and well-conditioned, i.e., its eigenvalues increase with time $T$. For example, we can choose a $\tau>0$ such that $\Gamma_{\tau}\succ 0$. Then, by monotonicity we have  $\sum^{T}_{t=0}\Gamma_t\succeq (T-\tau)\Gamma_{\tau}$.

The main technical difficulty is to control the difference between the Gram matrix and its expectation $\norm{V_T-\E V_T}$. Such a task might be possible in the case of strictly stable systems $\rho(\As)<1$, by using concentration inequalities and mixing arguments. However, this approach gives sample complexity bounds that explode as $\rho(\As)$ approaches $1$: two-sided concentration necessitates stability. Instead, we appeal to \emph{small-ball} techniques~\cite{mendelson2014learning}. Rather than bounding the difference between $V_T$ and its expectation, we only seek to obtain a one-sided lower bound.
The name small-ball refers to the fact that the distribution of $\lambda_{\min}(V_T)/T$ is not concentrated in a neighborhood of the origin---it exhibits anti-concentration. 

Define the extended covariance matrix
\[
\tilde{\Gamma}_{t}\triangleq\E \begin{bmatrix} x_t \\ u_t \end{bmatrix}\begin{bmatrix} x_t^\top & u_t^\top \end{bmatrix}  = \begin{bmatrix}
\Gamma_{t}&0\\0&\sigma^2_u I
\end{bmatrix}.
\]
Choose a time index $\tau>0$.
Invoking the small-ball methods described in the sidebar~\nameref{sidebar-small-ball}, it is possible to show that with probability at least $1-\delta$
\begin{equation}\label{ID_eq:PE_linear_systems}
V_T\succeq c\tau \left\lfloor \frac{T}{\tau }\right\rfloor\tilde{\Gamma}_{\lfloor \tau/2 \rfloor},
\end{equation}
where $c$ is universal constant,
provided that we have a large enough number of samples
\begin{equation}\label{ID_eq:Burn_In}
    T\ge \tau  O\Big((\dx+\du)\log \frac{\dx+\du}{\delta}+\log\frac{\det \tilde{\Gamma}_T}{\det \tilde{\Gamma}_{\lfloor \tau/2 \rfloor}}\Big).
\end{equation}
The right-hand side of the equation above increases with $T$; fortunately, under the assumption that the system is non-explosive $\rho(A)\le 1$, it increases at most logarithmically with $T$. Hence, condition~\eqref{ID_eq:Burn_In} will be satisfied for non-explosive systems for large enough $T$. The minimum time such that condition~\eqref{ID_eq:Burn_In} is satisfied is also known as the \emph{burn-in time}.

The time index $\tau$ gives us some control on the size of the lower bound $\tilde{\Gamma}_{\lfloor \tau/2 \rfloor}$. Recall that the sequence $\Gamma_t$ is increasing in the positive semi-definite cone. Hence, choosing a larger time index $\tau$, allows us to guarantee a stronger lower bound $\tilde{\Gamma}_{\lfloor \tau/2\rfloor}$. On the other hand, the required burn-in time increases linearly with $\tau$.

\begin{sidebar}{Persistency of excitation and small-ball bounds}
\section[Persistency of excitation and small-ball bounds]{}\phantomsection
   \label{sidebar-small-ball}
\setcounter{sequation}{0}
\renewcommand{\thesequation}{S\arabic{sequation}}
\setcounter{stable}{0}
\renewcommand{\thestable}{S\arabic{stable}}
\setcounter{sfigure}{0}
\renewcommand{\thesfigure}{S\arabic{sfigure}}

\sdbarinitial{L}et $z_t\in\Real^{\dz}$, $t\ge 0$ be a stochastic process adapted to a filtration $\{\F_t\}^{\infty}_{t=0}$. Let the Gram matrix be
\[
V_T=\sum_{t=0}^{T}z_tz'_t.
\]
We say that the process $z_t$ is persistently exciting with probability at least $1-\delta$ if there exist $c,\, T_0(\delta)>0$ such that
\[
\Prob(V_T\succeq cT I)\ge 1-\delta,
\]
for all $T\ge T_0(\delta)$. To prove persistency of excitation, we only need to establish one-sided lower bounds of the form
\[
\Prob(\lambda_{\min}(V_T)\ge cT)\ge1- \delta.
\]
In other words, we need to show that the least singular value of the Gram matrix does not concentrate in a small ball around the origin. We now discuss a sufficient condition first presented in~\cite{simchowitz2018learning} based on the small-ball method~\cite{mendelson2014learning}. An alternative approach via exponential inequalities can be found in \cite{ziemann2022note}.

\section{Block martingale small-ball condition}
Before establishing persistency of excitation for the whole vector $z_t$, we first study the projected processes $\xi^\top z_t$, where $\xi\in\Real^{\dz}$ is a unit vector. We say that the process $z_t$ satisfies the block martingale small-ball (BMSB) condition with parameters $(k,\Gamma_{\mathrm{lb}},p)$ if for every unit $\xi\in\Real^{\dz}$ and every $t\ge 0$
\begin{sequation}\label{ID_eq:block_martingale_small_ball_condition}
\frac{1}{k}\sum^k_{i=1}\Prob(|\xi^\top z_{t+i}|^2\ge \xi^\top \Gamma_{\mathrm{lb}} \xi|\F_{t})\ge p\text{ almost surely}.
\end{sequation}The above condition states that, conditioned on $t$, the block-average probability of being away from the origin is non-zero. The average probability is taken over blocks of size $k$.
The geometry of the lower bound is captured by the matrix $\Gamma_{\mathrm{lb}}$.

Let condition~\eqref{ID_eq:block_martingale_small_ball_condition} hold. Then, it follows that $z_t$ is persistently exciting, with the lower bound depending on the parameter $\Gamma_{\mathrm{lb}}$
\begin{sequation}\label{ID_eq:vector_small_ball}
    \Prob\big(V_T\succeq \frac{p^2}{16}k\lfloor T/k \rfloor \Gamma_{\mathrm{lb}}\big) \ge 1-\delta
\end{sequation}
as long as we have a large enough number of samples
\[
T\ge T_0= \frac{10k}{p}\left(\log \frac{1}{\delta}+2\dz \log \frac{10}{p}+\log\det (\Gamma_{\mathrm{ub}} \Gamma_{\mathrm{lb}}^{-1}) \right),
\]
with $\Gamma_{\mathrm{ub}}=\frac{\dz}{\delta} \max_{t\le T}\{\E z_tz_t'\}$.
Informally, the term $\Gamma_{\mathrm{ub}}$ is an upper bound of $V_T/T$, while the term $\Gamma_{\mathrm{lb}}$ is a lower bound of $V_T/T$. Hence the burn-in time $N_0$ depends logarithmically on the condition number of $V_T$.
The proof of the result can be found in~\cite{matni2019tutorial,simchowitz2018learning}.  

\section{Linear Systems}
In the case of fully-observed linear systems, we can select $z_t=\begin{bmatrix}x^\top_t&u^\top_t \end{bmatrix}^\top$ to be the vector of stacked state and input. Under white noise inputs, it can be shown~\cite{simchowitz2018learning} that the process $z_t$ satisfies the $(k,\tilde{\Gamma}_{\lfloor k/2\rfloor},3/20)$ block martingale small-ball condition, where
\[
\tilde{\Gamma}_{t}\triangleq \begin{bmatrix}
\Gamma_{t}&0\\0&\sigma^2_u I
\end{bmatrix}.
\]
\end{sidebar}

\subsubsection{Self-normalized term}
We begin with two observations about the self-normalized term 
\[
S_TV^{-1/2}_T=\left(\sum_{t=0}^{T-1} w_{t} \begin{bmatrix} x_t \\ u_t \end{bmatrix}^T \right)\left( \sum_{t=0}^{T-1} \begin{bmatrix} x_t \\ u_t \end{bmatrix}\begin{bmatrix} x_t \\ u_t \end{bmatrix}^T \right)^{-1/2}.
\]
First, we note that the process noise $w_t$ is independent of $x_t,\,u_t$ for all $t\le T$, i.e., the sum $S_T$ has a martingale structure. Second, as its name suggests, the term is self-normalized: if the covariates $x_t,\,u_t$ are large for some $t$, then any increase in $S_T$ will be compensated by an increase in $V^{-1/2}_T$. For this reason $S_TV^{-1/2}_{T}$ is called a \emph{self-normalized martingale}. Such terms have been studied previously in statistics in the asymptotic regime~\cite{lai1983asymptotic}. Here, we are interested in establishing finite sample bounds. We will invoke the results of~\citet{abbasi2011improved}---see the sidebar on~\nameref{sidebar-self-normalized} for more details.  Let $V$ be a symmetric positive definite matrix (\rev{to be decided later}) and set $\bar{V}_t=V_t+V$. \rev{The extra term $V$ guarantees positive definiteness of matrix $\bar{V}_t$.} Then
\begin{equation}\label{ID_eq:self_normalized_term}
\opnorm{S_T\bar{V}^{-1/2}_T}^2\le 8 \opnorm{\Sigmaw}\log\left(\frac{\det (\bar{V}_T)^{1/2}}{\det (V)^{1/2}} \frac{ 5^{\dx}}{\delta}\right).
\end{equation}
Crucially, self-normalization implies that the above term increases slowly (at most logarithmically) with the norm of $\bar{V}_T$. If the data is generated by a stable system, this dependency can be further reduced to order constant in the inverse stability margin \cite[see e.g.][Section 5.2]{jedra2020finite}.

In order to apply equation~\eqref{ID_eq:self_normalized_term}, we need to carefully select $V$. Moreover, to obtain data-independent sample complexity guarantees we require a data-independent upper bound of $\bar{V}_T$.
For the former, we choose $V=c\tau \left\lfloor \frac{T}{\tau }\right\rfloor\tilde{\Gamma}_{\lfloor \tau/2 \rfloor}$. When lower bound~\eqref{ID_eq:PE_linear_systems} on $V_T$ holds, we then also have that
 \[
\opnorm{S_TV^{-1/2}_T}^2\le 2\opnorm{S_T\bar{V}^{-1/2}_T}^2.
 \]
\rev{ For the latter, we may appeal to the matrix version of Markov's inequality (due to \citet[Theorem 12]{ahlswede2002strong}):}
\[
\Prob(V_T\not\preceq \frac{\dx+\du}{\delta}T\tilde{\Gamma}_T)\le \delta
\]
\rev{where $\{V_T\not\preceq \frac{\dx+\du}{\delta}T\tilde{\Gamma}_T\}$ is the complement of $\{V_T\preceq \frac{\dx+\du}{\delta}T\tilde{\Gamma}_T\}$.} Note that the application of Markov's inequality here is not particularly sub-optimal since $\bar{V}_T$ (and a factor $1/\delta$) already appears inside the logarithm in \eqref{ID_eq:self_normalized_term}.

\begin{sidebar}{Self-Normalized Martingales}
\section[Self-Normalized Martingales]{}\phantomsection
   \label{sidebar-self-normalized}
\setcounter{sequation}{0}
\renewcommand{\thesequation}{S\arabic{sequation}}
\setcounter{stable}{0}
\renewcommand{\thestable}{S\arabic{stable}}
\setcounter{sfigure}{0}
\renewcommand{\thesfigure}{S\arabic{sfigure}}

\sdbarinitial{A}n object that arises often in standard least squares analyses is the so called self-normalized martingale. Let $\{\F_t\}^{\infty}_{t=0}$ be a filtration and let $z_t\in\Real^{\dz}$, for some $\dz>0$, be a stochastic process such that $z_t$ is $\F_{t-1}$-measurable. Let $\eta_t\in\Real^{\dhh}$, $\dhh>0$, be a martingale difference sequence with respect to $\F_t$, i.e., $\eta_t$ is integrable, $\F_t$-measurable, with $\E(\eta_t|\F_{t-1})=0$. Then, a self-normalized martingale $M_k\in\Real^{\dhh\times\dz}$ is defined as
\[
M_{k}=\left(\sum_{t=0}^{k}\eta_t z^\top_t \right)\left(V+\sum_{t=0}^{k}z_t z^\top_t \right)^{-1/2},
\]
where $V$ is an arbitrary symmetric positive definite matrix of appropriate dimensions.

\section{Bounds for scalar processes}
Assume that $\eta_t\in\Real$ is a scalar process. Under some regularity conditions on the tail of $\eta_t$, we can establish finite sample bounds on the magnitude of $M_k$. Let the process $\eta_t$ be conditionally $K$-\emph{sub-Gaussian} for some $K>0$:
\[
\E (e^{\lambda \eta_t}|\F_{t-1})\le e^{\frac{K^2\lambda^2}{2}},\text{ for all }\lambda\in\Real.
\]
The above condition requires that the tails of $\eta_t$ decay at least as quickly as a Gaussian distribution. Now, we can invoke Theorem~1 of~\cite{abbasi2011improved}. 
Letting 
\[\bar{V}_k=V+\left(\sum_{t=0}^{k}z_t z^\top_t \right),\]
we then have the following finite sample bound. Pick a failure probability $\delta\in(0,1)$: then with probability at least $1-\delta$
\begin{sequation}\label{ID_eq:MART_scalar}
    \norm{M_k}_2^2\le 2K^2\log\left(\frac{\det (\bar{V}_k)^{1/2} }{\det (V)^{1/2}}\frac{1 }{\delta} \right)
\end{sequation}

\section{Extension to Vector Processes}
Assume now that the  process $\eta_t$ is \emph{vectored-valued}, with $\dhh>1$, and conditionally $K$-sub-Gaussian, i.e., for any unit vector $v\in\Real^{{\dhh}}$, $\norm{v}_2=1$, the projected process $v^\top \eta_t$ is conditionally $K$-sub-Gaussian. 
The bound~\eqref{ID_eq:MART_scalar} does not apply directly since it relies on the process $\eta_t$ being scalar. Nevertheless, 
by appealing to \emph{covering techniques}~\cite{vershynin2018high}, it is straightforward to generalize this argument to vector processes. 
The idea is to apply~\eqref{ID_eq:MART_scalar} to projections $v^\top \eta_t$ of $\eta_t$ onto several directions $v$ of the unit sphere. 

In particular, we discretize the unit sphere by considering points $v_i$, $i=1,\dots,N_\varepsilon$ such that the points are an $\varepsilon-$net, i.e., they cover the whole sphere with $\varepsilon-$balls around them. 
Then by taking a union bound over all points $v_j$, we obtain that with probability at least $1-\delta$
\begin{sequation}\label{ID_eq:MART_vector}
    \opnorm{M_k}^2\le 2(1-\varepsilon)^{-2}K^2\log\left(\frac{\det (\bar{V}_k)^{1/2}}{\det (V)^{1/2}} \frac{ N_{\varepsilon}}{\delta}\right),
\end{sequation}where the number of points is at most
\[
N_{\varepsilon}\le (1+\frac{2}{\varepsilon})^{\dhh}.
\]
The term $(1-\varepsilon)^{-2}$ comes from the discretization error and decreases as the discretization becomes finer. However, as the discretization becomes finer, the number of points $N_\varepsilon$ increases.  A typical choice is $\varepsilon=1/2.$

The above guarantees are with respect to the operator norm. We could also obtain guarantees for the Frobenius norm by applying~\eqref{ID_eq:MART_scalar} to $e^\top_i v_t$, where $e_i$, $i=1,\dots,\dhh$ are the canonical vectors of $\Real^{\dhh}$: in this case, with probability at least $1-\delta$
\begin{sequation}\label{ID_eq:MART_vector_frobenius}
    \norm{M_k}_F^2\le 2\dhh K^2\log\left(\frac{\det (\bar{V}_k)^{1/2}}{\det (V)^{1/2}} \frac{ \dhh}{\delta}\right).
\end{sequation}
\end{sidebar}

\begin{figure}[h]
 \centering
 \begin{subfigure}{0.5\textwidth}
\centering
\includegraphics[scale=0.5]{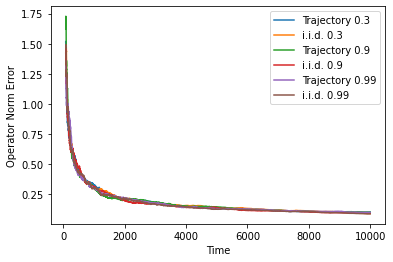}
\caption{We plot the operator norm error of least squares identification for $\rho(A_\star) \in \{0.3,0,9,0.99\}$, $\lambda_{\min}(A_\star)\approx 0$ and $\dx=25$. Lines marked "Trajectory" are sampled from a linear dynamical system $x_{t+1}=A_\star x_t +w_t$ whereas lines marked "i.i.d." are drawn from an independent baseline motivated by \cite{tu2022learning}. These i.i.d. lines correspond to a linear regression model $y_t = A_\star x_t +w_t$ in which the $x_t$ are drawn i.i.d. from $\mathcal{N}(0,\mathsf{dlyap}(A_\star,I_{\dx}))$. }
 \end{subfigure}
 \hfill
\begin{subfigure}{0.5\textwidth}
\centering
\includegraphics[scale=0.5,bb=0 0 14cm 10cm]{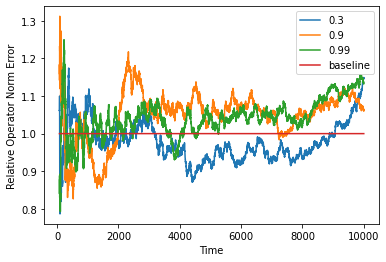}
\caption{Even as the correlation length $1/(1-\rho(A_\star))$ increases, the relative performance of the dynamic model to the independent baseline oscillates around $1$.}
\end{subfigure}
\caption{The plot shows the essence of the learning without mixing phenomenon \cite{simchowitz2018learning}: dependence does not necessarily impede the rate of convergence.}
\label{fig:lwom}
\end{figure}
\subsubsection{Sample Complexity Upper Bounds}
\label{sec:samplecomplexityub}
 Combining the previous bounds we finally obtain instance specific sample complexity upper bounds. For the least-squares estimator \eqref{ID_eq:LSEinAC}, we have that
 \begin{equation}\label{ID_eq:sample_complexity_ls_op_norm}
 \Prob(\opnorm{\thetas-\widehat\theta}\ge \varepsilon)\le \delta
 \end{equation}
 if the burn-in time condition~\eqref{ID_eq:Burn_In} is satisfied along with
 \begin{multline}\label{ID_eq:sample_complexity_ls_identification_error}
     T\ge c'\frac{\opnorm{\Sigmaw}}{\varepsilon^2 \lambda_{\min}(\Gamma_{\lfloor \tau/2 \rfloor})}\left( (\dx+\du) \log \frac{\dx+\du}{\delta}\right.\\
     \left.+\log\frac{\det \tilde{\Gamma}_T}{\det \tilde{\Gamma}_{\lfloor \tau/2 \rfloor}}\right), 
 \end{multline}
 where $c'$ is a universal constant. Once again the right-hand side of inequality~\eqref{ID_eq:sample_complexity_ls_identification_error} increases at most logarithmically with the estimation horizon $T$ for non-explosive systems ($\rho(A)\le 1$), and hence will be satisfied for large enough $T$. In fact, the rate defined in~\eqref{ID_eq:sample_complexity_ls_identification_error} is near-optimal in the sense that it nearly matches the linear regression rate achieved \emph{when all the samples are drawn independently}. See Figure~\ref{fig:lwom} for an illustration.

 To simplify the presentation, assume for now that we have strict stability $\rho(A)<1$. In this case the burn-in condition~\eqref{ID_eq:Burn_In} and sample complexity bound~\eqref{ID_eq:sample_complexity_ls_identification_error} can be combined and rewritten as
 \begin{equation*}
 T\ge c''\max\{\tau,\frac{1}{\varepsilon^2 \mathrm{snr}_\tau }\}(\dx+\du)\log \frac{\dx+\du}{\delta},
 \end{equation*}
 where $c''$ is another universal constant, and 
 \[\mathrm{snr}_\tau=\frac{\lambda_{\min}(\tilde{\Gamma}_{\lfloor \tau/2 \rfloor})}{\opnorm{\Sigmaw}}\]
 captures the ``signal to noise ratio'' of the system. The larger the $\mathrm{snr}$ the larger the excitation of the state compared to the magnitude of the noise. If the system has eigenvalues on the unit circle $(\rho(A)=1)$, then the expression looks similar but with some additional logarithmic terms; for simplicity, we omit this discussion here.

Ignoring logarithmic terms, the sample complexity grows as fast as $1/\varepsilon^2$, as we require more accuracy. Alternatively, the identification error decays as fast as $\tilde{O}(1/\sqrt{T})$, with the number of samples $T$. It also increases linearly with the dimension of the unknowns $\dx+\du$. Intuitively, matrices $\As$, $\Bs$ have $\dx^2+\dx\du$ unknown entries. Every state measurement has $\dx$ entries. Hence, we need at least $\dx+\du$ state samples to match the number of unknowns in $\As,\,\Bs$.  The sample complexity is also inversely proportional to the signal-to-noise ratio. Finally, it depends logarithmically on $\delta$, as (heuristically) predicted by the Central Limit Theorem.

It is worth mentioning that the signal to noise ratio depends heavily on the controllability structure of the system. In particular, under white-noise inputs, the state-covariance matrix $\Gamma_k$ is actually the controllability Gramian of the pair $(A,\begin{bmatrix} \sigma_u^2 B& \Sigmaw^{1/2} \end{bmatrix})$. In this setting, controllability is equivalent to excitability of the system. When the noise is isotropic (or non-singular), the noise covariance $\Sigmaw$ has full rank. Then, we can confirm that $\Gamma_1\succeq \Sigmaw\succ 0$, which implies that the state is directly excited. It is, thus, sufficient to select $\tau=2$ in the burn-in time condition~\eqref{ID_eq:Burn_In} and sample complexity bound~\eqref{ID_eq:sample_complexity_ls_identification_error}. When the noise is rank-deficient, the state can only be indirectly excited; \rev{we can still achieve persistency of excitation if there exists a $\tau>0$ such that $\tilde{\Gamma}_{\lfloor \tau/2 \rfloor}$ is non-zero. In particular, we can select $\lfloor \tau/2 \rfloor$ to be equal to the controllability index of the system~\cite{tsiamis2021linear}, that is the smallest possible $\kappa>0$ such that $\Gamma_{\kappa}\succ 0$.}

The above sample complexity upper bound is instance specific, i.e., it holds for a specific system $(\As,\Bs,\Sigmaw)$. To obtain class-specific sample complexity upper bounds for some class $\CC$, we need to impose global bounds on the norms of all $(\As,\Bs,\Sigmaw)\in \CC$ as well as a global bound on $\lambda^{-1}_{\max}(\Gamma_\tau)$, for some $\tau>0$--see for example~\cite{tsiamis2021linear}.

\subsubsection{Confidence ellipsoids}\label{sec:ellipsoids}
Sample complexity guarantees are qualitative and data-independent. That is, they provide intuition about how the number of required samples depends on various control theoretic parameters such as the dimension of the system, the signal to noise ratio, etc. These guarantees depend directly on the quantities of the unknown system being estimated---see equations~\eqref{ID_eq:Burn_In} and~\eqref{ID_eq:sample_complexity_ls_identification_error}---limiting their practical applicability. Another limitation is that the operator norm $\opnorm{\thetas-\widehat\theta}$ picks up the direction of largest error. As a result, a guarantee as in equations~\eqref{ID_eq:sample_complexity_ls_op_norm} and~\eqref{ID_eq:sample_complexity_ls_identification_error} provides confidence balls which can be conservative in certain directions of the state-space.

In practice, it might be more useful provide data-dependent confidence ellipsoids.
Towards this end, we can still apply the tools for self-normalized martingales presented in sidebar~\nameref{sidebar-self-normalized}. 
Let $V$ be symmetric positive definite and define $\bar{V}_t=V_t+V$. Using the properties of the least-squares estimator
\[
\opnorm{(\thetas-\widehat\theta)\bar{V}^{1/2}_T}^2\le \opnorm{S_T\bar{V}^{-1/2}_T}^2\opnorm{\bar V^{1/2}_TV_T^{-1/2}}^2.
\]
Define the ellipsoid radius to be
\[
r(\delta)\triangleq 8 \opnorm{\Sigmaw}\log\left(\frac{\det (\bar{V}_T)^{1/2}}{\det (V)^{1/2}} \frac{ 5^{\dx}}{\delta}\right)\opnorm{\bar V^{1/2}_TV_T^{-1/2}}^2.
\]
Invoking equation~\eqref{ID_eq:self_normalized_term}, we obtain
\begin{equation}\label{ID_eq:confidence_ellipsoids}
\Prob(\opnorm{(\thetas-\widehat\theta)\bar{V}^{1/2}_T}^2 \le r(\delta) )\ge 1-\delta.
\end{equation}
Interestingly, the ellipsoid adapts to the informativity of the data, as captured by $\bar{V}_T$. If some mode of the system is well-excited in $V_T$, the respective parameter error will be small.
With the exception of $\opnorm{\Sigmaw}$, all other quantities can be computed directly from data. In practice, one could replace $\opnorm{\Sigmaw}$ by an upper-bound or compute an empirical covariance from data. Although this quantity provides sharper confidence ellipsoids, it does not reveal directly how the identification error depends on the number of samples, i.e., it does not reveal the statistical rate of estimating $\thetas$. Other data-dependent methods for establishing confidence ellipsoids can be found in~\cite{dean2020_on_the_sample,care2018finite, matni2019tutorial}.

\subsubsection{Sample Complexity Lower Bounds}
The upper bounds on the sample complexity of system identification of the previous section are  only valid for the least squares estimator~\eqref{ID_eq:LSEinAC}. One may naturally ask whether we can do better with a different algorithm, i.e., are the sample requirements of the least squares algorithm a fundamental limitation or are they suboptimal? 
One way to answer these questions is by establishing minimax lower-bounds.
The main technical workhorse underpinning such lower bounds are information theoretic inequalities. 

As we will show next, the least squares identification algorithm analyzed above is nearly-optimal in the case of fully-observed systems.
To prove this, it is sufficient to construct system instances that are difficult to identify \emph{for all possible identification algorithms}. By invoking information theoretic inequalities, we can show that any algorithm requires at least as many samples as the least squares algorithm. 

\rev{We establish lower bounds for systems without exogenous inputs but the same results also apply to systems with white-noise exogenous inputs. For simplicity, we focus on the former case. Since there is no control input to implement an exploration policy, we denote this setting by $\pi=\emptyset$. Note that the case of more general exploration policies is an active front of research and is also discussed later on.} 
Fix a spectral radius $\rho$, and define the class of scaled orthogonal systems
\[
\mathcal{O}_\rho=\set{\As\in\Real^{\dx\times \dx}:\:\As=\rho O,\,O^\top O=I}.
\]
Let $N_c=N_c(\varepsilon,\delta, \mathcal{O}_\rho,\mathcal{A},\emptyset)$ denote the best possible sample complexity for learning over the class of scaled orthogonal systems. In~\cite{simchowitz2018learning}, it is shown that for any identification algorithm $\mathcal{A}$
\[
N_c= \Omega\left(\frac{\dx+\log 1/\delta}{\varepsilon^2\snr_{N_c}}\right).
\]
The result follows from a standard application of information theoretic lower bounds---see the sidebar on~\nameref{sidebar-Birge} for more details.
This shows that the rate $1/\varepsilon^2$, the dimension factor $\dx$, and the confidence $\log 1/\delta$ are fundamental, implying that the least-squares algorithm is near optimal.

 The above result holds for the specific subclass $\mathcal{O}_\rho$ of autonomous scaled orthogonal systems.
It is also possible to obtain stronger, \textbf{instance-specific} lower bounds, namely, lower bounds that hold locally around any fixed system. 
In particular, let $\thetas$ be an unknown system and consider a ball
$\mathcal{B}(\thetas,3\varepsilon)$ of radius $3\varepsilon$ around $\thetas$.
Let $N_c=N_c(\varepsilon,\delta, \mathcal{B}(\thetas,3\varepsilon),\mathcal{A},\emptyset)$ denote the minimum number of samples for identifying the local class $\mathcal{B}(\thetas,3\varepsilon)$.
In~\cite{jedra2022finite} it is shown that for any identification algorithm $\mathcal{A}$, and any failure probability $\delta\in(0,1)$ and accuracy
 $\varepsilon \in (0,\infty)$ it holds true that:
\[
N_c =\Omega\left(\frac{\dx+\log 1/\delta}{\varepsilon^2\snr_{N_c}}\right).
\]
The proof is also based on~\nameref{sidebar-Birge}.

Terms capturing the $\snr$, appear in both upper and lower bounds. However, there is a gap between the upper and lower bounds. The former depend on $\lambda^{-1}_{\min}(\Gamma_\tau)$, for some small enough $\tau$, while the latter depend on $\lambda^{-1}_{\min}(\Gamma_T)$, where $T$ is the number of samples collected. Note that we cannot increase $\tau$ too much, since it affects the burn-in time condition~\eqref{ID_eq:Burn_In}.
In the case of stable systems $\rho(\As)<1$, this gap can be closed 
at the expense of a burn-in time that depends on the mixing time $1/(1-\rho(\As))$ of the system ~\cite{jedra2020finite}. The gap can be also made small, i.e., $\tau=\Theta(T)$, in the case of diagonalizable marginally stable systems with $\rho(\As)=1$~\cite{tu2022learning}. 

In the case of systems with white-noise control inputs the same analysis can be applied. In the case of general exploration policies the landscape is more complex, since both the policy $\pi$ and the identification algorithm $\mathcal{A}$ affect sample complexity. Let $N_c=N_c(\varepsilon,\delta, \mathcal{B}(\thetas,3\varepsilon),\mathcal{A},\pi)$ be the local sample complexity defined as before, where now the policy $\pi$ can also be varied. Following the result of~\cite{jedra2019sample}, we obtain the lower bound condition
\[
N_c= \Omega\left(\frac{\log 1/\delta}{\varepsilon^2\snr^\star_{N_c}}\right),
\]
where the exploration policy $\pi$ is chosen to optimize the $\snr$ term:
\[
\snr^\star_{N_c}=\max_{\pi}\frac{1}{N_c}\sum_{t=1}^{N_c}\E\begin{bmatrix} x_t \\ u_t \end{bmatrix}\begin{bmatrix} x_t^\top & u_t^\top \end{bmatrix}. 
\]
In order to avoid arbitrarily large exploration inputs, we we limit the control input energy
\[ 
\E \norm{u_t}^2_2\le \sigma^2_u,
\]
for some $\sigma_u>0$, as otherwise, we trivially obtain $\snr^\star=\infty$. Finding the optimal exploration policy is not a simple problem and requires knowledge of the system dynamics. In~\cite{wagenmaker2020active} it is shown that the above lower bound can be achieved asymptotically (as $\delta\rightarrow 0$) by following an active exploration policy based on sinusoidal signals. 

\begin{sidebar}{Birgé's Inequality}
\section[Birgé's Inequality]{}\phantomsection
   \label{sidebar-Birge}
\setcounter{sequation}{0}
\renewcommand{\thesequation}{S\arabic{sequation}}
\setcounter{stable}{0}
\renewcommand{\thestable}{S\arabic{stable}}
\setcounter{sfigure}{0}
\renewcommand{\thesfigure}{S\arabic{sfigure}}

\sdbarinitial{B}irgé's inequality  is a sharper version of Fano's inequality, a classical tool from information theory \cite{boucheron2013concentration}. It can be used to establish lower bounds in multiple testing problems. Before we state the inequality, recall the definition of Kullback–Leibler (KL) divergence between two probability distributions $(\Prob, \mathbf{Q}),$
\[
D(\mathbf{Q}||\Prob)\triangleq \E_{\mathbf{Q}}(\log \frac{d\mathbf{Q}}{d\Prob}),
\]
where we assume that $\mathbf{Q}$ is absolutely continuous with respect to $\Prob$ and $\frac{d\mathbf{Q}}{d\Prob}$ denotes the density of $\mathbf{Q}$ with respect to $\Prob$.
Now let $\Prob_0,\dots,\Prob_n$ be probability distributions over some measurable space $(\Omega,\F)$, such that $\Prob_i$, $i=1,\dots,n$ are absolutely continuous with respect to $\Prob_0$. These probability distributions represent, for instance, different hypotheses in a multiple hypothesis testing scenario.
Let $E_0,\dots,E_n\in \F$ be disjoint events. For instance, $\Prob_i(E_i)$ might represent the probability of making a correct guess.
Birgé's inequality states that a necessary condition for the minimum success-probability to be lower bounded as
\begin{sequation}\label{ID_eq:Birge_probabilities}
\min_{i=0,\dots,n} \Prob_{i}(E_i)\triangleq 1-\delta \ge \frac{1}{n+1}.
\end{sequation}
is that the average pairwise KL divergence between the $\Prob_i$ and $\Prob_0$ satisfies the lower bound
\begin{sequation}\label{ID_eq:Birge}
    \frac{1}{n}\sum_{i=1}^{n}D(\Prob_i||\Prob_0)\ge h(1-\delta,\delta/n),
\end{sequation}where $h(p,q)=p\log p/q +(1-p)\log(1-p)/(1-q)$.
The above condition states that making a correct guess with high probability is possible \rev{only if the distributions $\Prob_1,\dots\Prob_n$ are sufficiently distinguishable from $\Prob_0$. Note that condition~\eqref{ID_eq:Birge_probabilities} is permutation invariant, i.e. it is independent of the ordering of the probability distributions. Hence, Birgé's inequality~\eqref{ID_eq:Birge} should also hold if we swap $\Prob_0$ with any $\Prob_j$, $j\le n$. Hence $\Prob_0,\dots,\Prob_n$ should be mutually distinguishable.}

\section{System Identification}
Let $\CC=\{\theta_0,\dots,\theta_n\}$ be a class of systems that are $2\varepsilon$-separated, i.e., $\norm{\theta_i-\theta_j}> 2\varepsilon$. Let $\Prob_i$ be the probability distribution of the data $\{(y_0,u_0),\dots,(y_T,u_T)\}$ when the underlying system is $\theta_i$. Let $\widehat{\theta}$ be the output of any identification algorithm. Since the systems are separated, the events $E_i\triangleq \{\norm{\theta_i-\widehat\theta}\le \varepsilon \}$ will be disjoint. If some algorithm performs well with high probability across all systems, then~\eqref{ID_eq:Birge_probabilities} holds, which, in turn, implies that~\eqref{ID_eq:Birge} holds.

To obtain the tightest lower bounds possible, we aim to construct sets of $2\varepsilon-$separated systems which nonetheless lead to data distributions with small KL divergence. In other words, the separation should not be too large, so that the distributions are as indistinguishable as possible.
\end{sidebar}

\begin{table*}
\renewcommand{\arraystretch}{1.5}
\caption{Sample Complexities of Fully-Observed System Identification.\textrm{ We define $\dd=\dx+\du$. The total number of non-zero elements is denoted by $\dsparse$. By $\snr^\star$, we denote the snr under the best possible active exploration policy. For~\cite{sarkar2019near}, we only show the result for $\rho(\As)\le 1$. The sample complexities are given in terms of $\Ntot=\Ntraj T$, i.e. the total number of samples, where $T$ is the horizon and $\Ntraj$ is the number of trajectories. For single trajectory data, we have $\Ntot=T$. All bounds are non-asymptotic and we only use the big-O notation to simplify the presentation of the bounds.} \label{ID_TAB:Sys_ID_Fully}}
\center{
\begin{tabular}{@{}|c|c|c|c|cc|c|c|@{}}
\hline
Paper&Trajectory&Stability& Actuation &Upper Bound&Burn-in time&Lower Bound\\
\hline
\cite{dean2020_on_the_sample} &multiple& any &white-noise & $\tilde O(T\frac{\dd\log 1/\delta}{\varepsilon^2\snr_{T} })$&$T\tilde O(\dd+\log1/\delta)$& - \\
\cite{simchowitz2018learning} &single&$\rho(\As)\le 1$ &white-noise & $\tilde O(\frac{\dd\log \dd/\delta}{\varepsilon^2 \snr_\tau })$ &$\tilde O(\tau\dd \log \dd/\delta)$ & $\Omega(\frac{\dd+\log 1/\delta}{\varepsilon^2 \snr_{T}})$\\
\cite{sarkar2019near}&single&any&white-noise&$\tilde O(\frac{\dd\log \dd/\delta}{\varepsilon^2 \snr_1 })$&$\tilde O(\dd \log \dd/\delta)$ &-
\\
\cite{jedra2019sample}& single&any& active &- & -&$\Omega(\frac{\log 1/\delta}{\varepsilon^2\snr^\star_T})$  \\
\cite{jedra2020finite} & single &$\rho(\As)< 1$&white-noise& $\tilde O(\frac{\dd+\log 1/\dd}{\varepsilon^2 \snr_T})$&$\tilde O(\frac{\dd+\log 1/\dd}{(1-\rho(\As))^2 
})$& -  \\
\cite{wagenmaker2020active} & single&$\rho(\As)<1$&active& $\tilde O(\frac{\dd+\log 1/\delta}{\varepsilon^2 \snr^\star_\tau })$&$ \poly(\frac{1}{1-\rho(\As)})\tilde O(\dd+\log 1/\delta)$&$\Omega(\frac{\log 1/\delta}{\varepsilon^2 \snr^\star_\infty})$\\
\cite{fattahi2019learning} & single& $\rho(\As)<1$ & white-noise & $\tilde O\big (\frac{\dsparse \log \dd/\delta}{\varepsilon^2 \snr_{\infty}(1-\rho(\As))}\big)$&$\tilde O\big(\frac{\dsparse^2 \log \dd/\delta}{ 
(1-\rho(\As))^4}\big)$ &-\\
\cite{tsiamis2021linear}&single&$\rho(\As)\le 1$&any&$ \tilde{O}(\exp(\dd)\frac{\log 1/\delta}{\varepsilon^2})$
&$\tilde{O}(\dd\log \dd/\delta)$ &$\Omega(\exp(\dd)\frac{\log 1/\delta}{\varepsilon^2})$\\
\hline
\end{tabular}
}
\end{table*}

\subsubsection{Summary and Generalizations}
In Table~\ref{ID_TAB:Sys_ID_Fully}, we summarize some of the main results for the sample complexity of identifying fully-observed systems. For compactness, we denote $\dd=\dx+\du$.
Only results for open-loop non-explosive systems $(\rho(\As)\le 1)$ are shown. 

If a stabilizing feedback gain $K_0$ is somehow known beforehand, the results can immediately be extended to the case of closed-loop stable systems $(\rho(\As-\Bs K_0)< 1)$ under the stabilizing feedback law $u_t=K_0x_t+\eta_t$. 
The case of open-loop unstable systems with $\rho(\As)>1$  is analyzed in~\cite{faradonbeh2018finite_ID,sarkar2019near}, where it is shown that under a regularity condition on the eigenvalues of $\As$, the error of learning explosive systems decays exponentially quickly with the number of samples. In~\cite{sarkar2019near} it is further shown that the error of learning systems with all eigenvalues on the unit circle
decays at least as fast as $\tilde O(1/T)$ as opposed to the $\tilde{O}(1/\sqrt{T})$ error we get for strictly stable systems.
The above rates agree with previous asymptotic results~\cite{lai1983asymptotic}. 

As we discussed in the presentation of the lower bounds, the least squares algorithm is near optimal in the case of white-noise excitation. In the case of non-explosive systems $\rho(\As)=1$, there is a gap between the upper and lower bounds.
The gap can be closed in the case of stable systems $\rho(\As)<1$~\cite{jedra2020finite}. This can be achieved by exploiting the~\nameref{sidebar-Hanson} (see sidebar for more details) instead of small-ball techniques. However, the downside of using Hanson-Wright is that the burn-in time depends on the mixing time of the system $1/(1-\rho(\As))$. As the system approaches instability $\rho(\As)\rightarrow 1$, then the finite sample guarantees degrade rapidly due to the burn-in time going to infinity. A benefit of small ball techniques is that they hold even in the regime $\rho(\As)=1$.

In the presentation of sample complexity upper bounds, we only considered white-noise input signals. Although white-noise input signals can guarantee persistency of excitation and lead to parameter recovery, they constitute a suboptimal exploration policy. It is a passive form of exploration that does not adapt online to the gathered information. Instead, in~\cite{wagenmaker2020active}, an active exploration policy is employed based on sinusoidal inputs, leading to sharper sample complexity guarantees. In fact, in the regime where the failure probability goes to zero $\delta\rightarrow 0$, the proposed active exploration policy together with the least squares identification algorithm are near-optimal and achieve the minimax lower bound. 

Another interesting problem is sparse system identification, where there might be an underlying sparse structure in the matrices $(\As,\Bs)$. In~\cite{fattahi2019learning}, it is shown that under an $\ell_1$-regularization penalty and certain mutual incoherence conditions, the sample complexity of correctly identifying the non-zero elements of $(\As,\Bs)$ scales with $\dsparse^2$, i.e., the number of non-zero elements, instead of the problem's dimensions $\dx+\du$. Hence, if the non-zero elements are fewer than the dimension of the problem, we suffer from a smaller sample complexity. It is an open problem whether the power of $\dsparse$ can be improved. Moreover, it is an open question whether the results can be extended to open-loop non-explosive systems $\rho(\As)=1$; currently, the burn-in time depends on the mixing time $1/(1-\rho(\As-K_0\Bs))$, where $K_0$ is a stabilizing gain, known a priori. 

So far, we have focused on single trajectory data. In practice, we might have access to data generated by several trajectories. 
In~\cite{dean2020_on_the_sample,tu2022learning}, learning from multiple independent trajectories is studied, where $\Ntot=\Ntraj T$ is the total number of samples, $T$ is the trajectory length, and $\Ntraj$ is the number of trajectories. In~\cite{dean2020_on_the_sample}, many samples are discarded (all but the last two) to turn system identification into an i.i.d. regression problem. As a result, there is an $O(T)$ extra sample overhead. These limitations are addressed by~\cite{tu2022learning}, where single trajectory and multiple trajectory learning were treated in a unified way; the parameter recovery guarantees are different and given in expectation, hence, we did not include them in Table~\ref{ID_TAB:Sys_ID_Fully}. An interesting conclusion in~\cite{tu2022learning} is that in the ``many" trajectories regime, e.g. $\Ntraj=\Omega(\dd)$, learning is more efficient that in the ``few" trajectories regime, e.g. $\Ntraj=o(\dd)$. Hence, it might be more beneficial to increase the number of trajectories $\Ntraj$ rather than the horizon $T$, while keeping the total number of samples constant.

All previous results rely on the process noise being full rank with positive definite covariance $\Sigmaw \succ 0$. In this case, all modes of the system are directly excited by the process noise, making learning easier, as the system $\snr$ is always lower bounded by the condition number of the noise, i.e., $\snr_t\ge \frac{\opnorm{\Sigmaw}}{\lambda_{\min}(\Sigmaw)}$. As a result, in this case, system identification exhibits sample complexity, which scales polynomially with the system dimension $\dd$. If we take away this structural assumption and allow degenerate noise, then, sample complexity can increase dramatically. In~\cite{tsiamis2021linear}, it is shown that there exist non-trivial classes of systems for which the sample complexity scales exponentially with the dimension $\dd$.
Such classes include underactuated systems, e.g. systems with integrator/network structure. Such systems are structurally hard to control/excite, and, thus, difficult to identify. Under an additional robust controllability requirement, it is shown in~\cite{tsiamis2021linear} that the sample complexity of identifying underactuated systems cannot be worse than exponential with the dimension $\dd$. In fact, it cannot be worse than exponential in the so called controllability index, which quantifies the degree of underactuation of a system.

\rev{Finally, we can obtain finite sample guarantees if the process noise sequence is a martingale difference sequence~\cite{simchowitz2020naive}, thus relaxing the i.i.d. requirement. Still, the methods presented here at quite fragile to the martingale difference noise assumption, which essentially amounts to a strong realizability assumption, implying in some sense that the model class contains the true model.  In certain situations with colored noise, it still possible to reduce the problem to a white noise problem---allowing us invoke the self-normalized martingale inequality---for instance by fitting a filter of sufficient length \cite{tsiamis2019finite}. However, in full generality, sharply dealing with colored noise in the non-asymptotic regime is very challenging.} 
\rev{If one seeks to go beyond sub-Gaussian tails the situation becomes even more subtle. In a heavy-tailed noise model, with for instance $\E\|w_t\|^4<\infty$ but $\E\|w_t\|^p=\infty$ for some finite $p>4$ then the least squares estimator is still optimal in expectation for most problems (at least for i.i.d. data \cite{mourtada2022exact}). However, it is no longer optimal in deviation---not even for i.i.d. data---meaning that it does not uniformly in $\delta$ attain the optimal $\log (1/\delta)$ failure probability \cite{oliveira2016lower}. Still for i.i.d. data, this optimal dependency can however we obtained by an alternative estimator (obtained by minimizing the so-called Huber loss, see \cite[Section 6.4]{mendelson2018learning}). We do not know of any results that sharply characterize the failure probability in heavy-tailed linear system-identification.}

\begin{sidebar}{The Hanson-Wright Inequality}
\section[Hanson-Wright Inequality]{}\phantomsection
   \label{sidebar-Hanson}
\setcounter{sequation}{0}
\renewcommand{\thesequation}{S\arabic{sequation}}
\setcounter{stable}{0}
\renewcommand{\thestable}{S\arabic{stable}}
\setcounter{sfigure}{0}
\renewcommand{\thesfigure}{S\arabic{sfigure}}

\sdbarinitial{I}n many situations of interest, e.g., when analyzing Gram matrices, we need to work with quadratic functions of random variables. The Hanson-Wright inequality~\cite{vershynin2018high} is a standard tool for analyzing concentration of such quadratic forms when the underlying random variables are sub-gaussian. Let $X=(X_1,\dots,X_n)\in\Real^n$ be a random vector with independent mean zero $K$-sub-gaussian coordinates satisfying
\[
\E e^{tX_i}\le e^{\frac{K^2t^2}{2}},\,i=1,\dots,n.
\]
Let $M\in\Real^{n\times n}$ be a matrix. Then, there exists a universal constant $c$ such that for every $s\ge 0$, we have

\begin{align*}
&\Prob(|X^\top M X-\E X^\top M X|\ge K^2s )\le 2 e^{-c\min\{\frac{s^2}{\norm{M}^2_F}, \frac{s}{\opnorm{M}}\}}.
\end{align*}
Hanson-Wright has been used as an alternative method for establishing persistency of excitation in the case of  identification of fully-observed, stable systems~\cite{jedra2020finite}. Contrary to small-ball methods, Hanson-Wright inequality is a two-sided result, which is a stronger requirement. Hence, it can be conservative in the case of unstable or marginally stable systems. Hanson-Wright inequality has also been utilized for proving ~\nameref{sidebar-Hankel_isometry} when the elements of the Hankel matrix are i.i.d.  
\end{sidebar}

\subsection{Partially Observed Systems}
We now consider the more general case of partially observed systems with $\Cs\neq I$ and $\Sigmav\neq 0$. Partial observability makes system identification harder as we do not have direct access to state measurements. In the case where we do not know anything about the system, identifying the ``true'' state-space parameters is impossible as the state-space representation is no longer unique, as the input-output map from inputs $u$ to measured outputs $y$ remains the same under similarity transformations. That is, for any invertible matrix $\Xi$, the following systems
\begin{align*}
\thetas&=(\As,\Bs,\Cs,\Sigmaw,\Sigmav)\\
\thetas'&=(\Xi^{-1}\As \Xi,\Xi^{-1}\Bs,\Cs \Xi,\Xi^{-1}\Sigmaw \Xi^{-\top},\Sigmav)
\end{align*}
are equivalent from an input-output point of view. Another source of ambiguity is that the noise model is also non-unique~\cite{van2012subspace}. 
Consider the system
\begin{equation}\label{ID_eq:KF_form}
\begin{aligned}
    \hat{x}_{k+1}&=\As\hat{x}_k+\Bs u_k+\Lk e_k\\
    y_{k}&=\Cs \hat{x}_{k}+e_k,
\end{aligned}
\end{equation}
where $\Lk$ is the steady-state Kalman filter gain
\begin{align*}
 \Lk&=\As \Ss\Cs^\top(\Cs\Ss\Cs^\top+\Sigmav)^{-1}\\
\Ss&=\As \Ss\As^\top+\Sigmaw-\As \Ss\Cs ^*(\Cs \Ss\Cs^\top+\Sigmav)^{-1}\Cs \Ss\As^\top.
\end{align*}
The innovation error is defined as
\[
e_k\triangleq y_k-\Cs\hat{x}_k.
\]
The innovation process is i.i.d., zero-mean Gaussian with covariance $\Sigmae\triangleq \Cs\Ss\Cs^\top+\Sigmav$~\cite{anderson2005optimal}.

System~\eqref{ID_eq:KF_form} is called the (steady-state) Kalman filter form or innovations form of system~\eqref{ID_eq:system}. Under the assumption that the system is initialized under its stationary distribution, i.e., that $\Gamma_0=\Ss$, system~\eqref{ID_eq:system} and its innovation form~\eqref{ID_eq:KF_form} are statistically equivalent from an input-output perspective in that they generate outputs with identical statistics. It has been common practice in the system identification literature~\cite{qin2006overview} to work with the representation~\eqref{ID_eq:KF_form} instead of the original system~\eqref{ID_eq:system}. One reason is that the innovation noise is always output-measurable, as opposed to the process/measurement noise. Another reason is that under certain observability conditions, the closed-loop map $\As-\Lk \Cs$ is stable, i.e., $\rho(\As-\Lk \Cs)<1$.

We present techniques which can be applied to open-loop non-explosive systems that satisfy $\rho(\As)\le 1$. We again assume that the open-loop inputs are white noise zero-mean Gaussian, i.i.d., with $\E u_t u^{\top}_t=\sigma^2_u I$, for some $\sigma_u>0$. We also assume that $(\As,\Cs)$ is detectable, $(\As,\Sigmaw^{1/2})$ is stabilizable, and $\Sigmav$ is invertible so that the innovation form~\eqref{ID_eq:KF_form} is well-defined and $\rho(\As-\Lk \Cs)<1$. To simplify the analysis, we assume that the Kalman filter starts from its steady-state $\Gamma_0=\Ss$, $\E x_0=0$. The latter is a weak assumption; due to the stability of the Kalman filter, we will converge to the steady-state exponentially fast.

Most identification methods
follow the prediction error approach~\cite{Ljung1999system} or the subspace
method~\cite{van2012subspace}. The prediction error approach is typically
non-convex and directly searches over the system parameters
$\thetas$ by minimizing a prediction error cost. In the
subspace approach, Hankel matrices
of the system are estimated first based on a convex regression problem. Then, realization is performed, typically based on Singular Value Decomposition (SVD). In this survey, we focus on the subspace/realization approach. Prior work on the analysis of the prediction error method can be found in~\cite{hardt2018gradient}.

\subsubsection{Regression Step}\label{ID_sec:Regression}
The first step is to establish a regression between future outputs and past inputs and outputs.  Let $p>0$ be a past horizon. By unrolling the innovation form~\eqref{ID_eq:KF_form}, at any time step $k>0$, we can express $y_k$ as a function of $p$-past outputs and inputs
\begin{equation}\label{ID_eq:PO_main_regression}
    y_k=\underbrace{\Cs \Contr_p}_{G_p} Z_k+\underbrace{\Cs(\As-\Lk \Cs)^p\hat{x}_{k-p}}_{\text{bias}}+e_k,
\end{equation}
where $Z_k$ is the vector of all regressors stacked:
\[
Z_k=\begin{bmatrix}y^\top_{k-1}&u^\top_{k-1}&\cdots&y^\top_{k-p}&u^\top_{k-p}\end{bmatrix}^\top,
\]
and $\Contr_p$ is an extended controllability matrix:
\[
\Contr_p\triangleq \begin{bmatrix}\begin{bmatrix}\Bs&\Lk\end{bmatrix}
&\cdots&(\As-\Lk\Cs)^{p-1}\begin{bmatrix}\Bs&\Lk\end{bmatrix}\end{bmatrix}.
\]
Equation~\eqref{ID_eq:PO_main_regression} shows that there is a linear relation between future outputs and past inputs/outputs, which is determined by matrix $G_p=\Cs\Contr_p$. We have a linear regression problem which is similar to the one encountered in the fully-observed case since the innovation process $e_t$ is i.i.d. and the regressors $Z_k$ are independent of $e_k$ at time $k$. The main differences are that i)  there exists a bias error term and ii) the unknown matrix $G_p$ has a special structure. We can deal with the bias by increasing the past horizon $p$; the bias term goes to zero exponentially fast due to the stability of the Kalman filter. 

The above step is common in both prediction error and subspace identification methods. In the prediction error approach, we optimize over the original state-space parameters, e.g. $A,B,C$ etc, hence preserving the special structure of $G_p$. In the subspace approach, we do not optimize over the original system parameters. Instead, we optimize directly over the higher-dimensional representation $G_p$ by treating it as an unknown without structure. This leads to a convex least-squares problem
\begin{equation}\label{ID_eq:PO_ls}
    \hat{G}_{p,T}\in  \argmin_{G} \sum_{t=p}^{T} \|y_{t}-GZ_t\|_2^2.
\end{equation}
In machine learning, this lifting to higher-dimensions is referred to as improper learning~\cite{kozdoba2019line}. After some algebraic manipulations, we can verify that
\[
\hat{G}_{p,T}-G_p=\left(\sum^T_{t=p}e_tZ^\top_t \right)\left(\sum^T_{t=p}Z_t Z^\top_t \right)^{-1}+\text{bias},
\]
where the bias terms includes factors $(\As-\Lk\Cs)^{p}$ which decay exponentially  with the past horizon $p$.

The analysis now proceeds in a similar way as in the case of fully-observed systems. We break the least squares error into two terms, a self-normalized term and a term capturing persistence of excitation:
\[
\opnorm{\hat{G}_{p,T}-G_p}\le \opnorm{S_{p,T} V^{-1/2}_{p,T}}\opnorm{V_{p,T}^{-1/2}},
\]
where $S_T$, and $V_T$ are analogously defined as
\[
S_{p,T}=\sum^T_{t=p}e_tZ^\top_t,\,V_{p,T}=\sum^T_{t=p}Z_tZ^\top_t.
\]
For the self-normalized term, we exploit the techniques for~\nameref{sidebar-self-normalized}. For the second term, we need to show persistency of excitation. One way is to use again the small-ball techniques discussed in the fully-observed case. An alternative way is  establishing~\nameref{sidebar-Hankel_isometry}.

Using the tools listed above, we can obtain sample complexity upper bounds for recovering the matrix $G_p$.
Let $\Gamma_{Z,k}=\E Z_kZ^{\top}_k$ be the covariance of the regressors. For example, in the case of no inputs $\Bs=0$, ~\citet{tsiamis2019finite} show that under the least-squares algorithm defined above we have that
\[
\Prob(\opnorm{G_p-\hat{G}_{p,T}}\ge \varepsilon)\le \delta
\]
if we select $p=\Omega(\log T)$ and 
\begin{align*}
T\ge c \frac{p}{\varepsilon^2 \snr_p}\dy\log\left(\frac{p\dy}{\delta}\frac{\opnorm{\Gamma_{Z,T}}}{\lambda_{\min}(\Gamma_{Z,p})}\right),
\end{align*}
where $c$ is a universal constant and the signal to noise ratio is defined as
\[
\snr_k=\frac{\opnorm{\Sigmae}}{\lambda_{\min}(\Gamma_{Z,p})}.
\]
When we have inputs $\Bs\neq 0$, we can obtain a similar result by repeating the same arguments as in~\cite{tsiamis2019finite} and replacing $\dy$ with $\dy+\du$.
Once again we recover a rate of $\tilde{O}(1/\varepsilon^2)$. Equivalently, the error scales as $\tilde{O}(1/\sqrt{T})$. The main caveat is that we need to select $p$ to increase logarithmically with the horizon $T$ \rev{to mitigate the bias term}.
Ignoring $\varepsilon$, the $\snr$ and other system-theoretic parameters, we obtain that the sample complexity upper bound scales with $p(\dy+\du)$, i.e., it depends at linearly on the size of the past horizon $p$. \rev{This upper bound suggests that there is a tradeoff between reducing the bias term (large $p$) and reducing sample complexity (small $p$).}
This dependence on the past horizon $p$ arises because we ignore the structure of $G_p$ and we treat it as an unknown matrix. In this case $G_p$ has  $p(\dy+\du)\dy$ unknown entries. Since every measurement $y_k$ contributes with $\dy$ components, then a sample complexity of $O(p(\dy+\du))$ suffices. However, it might be the case that this sample complexity is suboptimal since the true number of unknowns in $\thetas$ is of the order of $\dx^2+\dx(\dy+\du)$. It seems that by lifting the problem to higher dimensions in~\eqref{ID_eq:PO_ls}, we suffer from larger sample complexity.

\begin{sidebar}{Isometry for Hankel Matrices}
\section[Isometry for Hankel Matrices]{}\phantomsection
   \label{sidebar-Hankel_isometry}
\setcounter{sequation}{0}
\renewcommand{\thesequation}{S\arabic{sequation}}
\setcounter{stable}{0}
\renewcommand{\thestable}{S\arabic{stable}}
\setcounter{sfigure}{0}
\renewcommand{\thesfigure}{S\arabic{sfigure}}

\sdbarinitial{L}et $\eta_0,\dots,\eta_{N-1}$ be a sequence of i.i.d. zero-mean isotropic Gaussian variables in $\Real^{\dhh}$, that is $\eta_t\sim\mathcal{N}(0,I_{\dhh})$, and consider the following Hankel matrix
\[
H_{L,N}\triangleq \begin{bmatrix}
\eta_0&\eta_1&\cdots&\eta_{N-L-1}\\
\vdots\\
\eta_L&\eta_{L+1}&\cdots&\eta_{N-1}
\end{bmatrix}.
\]
Such matrices arise in the analysis of system identification algorithm that use information of the past $L$ steps for prediction. For example $\eta_t$ could be the input process $u_t$ and/or the (normalized) innovations $e_t$.
A crucial problem is determining whether the matrices $H_{L,N}$ are persistently exciting. One solution is to exploit the small-ball approach as reviewed in sidebar~\nameref{sidebar-small-ball}. 

Here, we will review an alternative way to answer this question, which leads to a stronger two-sided result~\cite{lee2022improved,djehiche2021finite}.  Fix a failure probability $\delta\le 1/2$. 
Then there exists a universal constant $c$ such that if
\[
N\ge c L\dhh \log\frac{L\dhh}{\delta},
\]
then with probability at least $1-\delta$
\[
\frac{N}{2}I_{L\dhh}\preceq H_{L,N}H^\top_{L,N} \preceq \frac{3N}{2}I_{L\dhh}.
\]
The result is adapted from Theorem~A.2 in~\cite{lee2022improved}. The proof is based on the~\nameref{sidebar-Hanson} along with Fourier domain techniques. Similar results appeared in~\cite{sarkar2021finite,oymak2021revisiting} but require slightly larger burn-in time.
\end{sidebar}

\subsubsection{Realization}
Let us introduce the notation $\Acls\triangleq \As-\Lk\Cs$ and $\Hs\triangleq \begin{bmatrix}\Bs&\Lk \end{bmatrix}$. For this section, assume for simplicity that system $(\Cs,\Acls,\Hs)$ is\textbf{ minimal}, i.e. $(\Cs,\Acls)$ is observable and $(\Acls,\Hs)$ is controllable. Under this notation, matrix $G_p$ contains the Markov parameters $\Cs \Acls^{k}\Hs$, $k\le p-1$ of system $(\Cs,\Acls,\Hs)$, allowing for the use of standard realization techniques to extract $(\Cs,\Acls,\Hs)$ from the Markov parameters.  A standard such approach is the Ho-Kalman realization technique. If we assume that we know the true Markov parameters $G_p$, then we can construct the following Hankel matrix
\[
\Hk_{k_1,p}\triangleq\begin{bmatrix} \Cs\Hs &\Cs\Acls\Hs&\cdots&\Cs\Acls^{p-1-k_1}\Hs\\\Cs\Acls\Hs&\Cs\Acls^2\Hs&\cdots&\Cs\Acls^{p-k_1}\Hs\\ \vdots &&\ddots&\vdots\\\Cs\Acls^{k_1}\Hs &\Cs\Acls^{k_1+1}\Hs&\cdots&\Cs\Acls^{p-1}\Hs\end{bmatrix}.
\]
The Hankel matrix has rank $\dx$, since it can be written as the outer-product of a controllability and an observability matrix:
\[
\Hk_{k_1,p}=\underbrace{\begin{bmatrix}\Cs\\\Cs\Acls\\\vdots\\ \Cs\Acls^{k_1}\end{bmatrix}}_{\obs_{k_1}}\underbrace{\begin{bmatrix}\Hs&\Acls\Hs&\cdots& \Acls^{p-1-k_1}\Hs\end{bmatrix}}_{\contr_{p-1-k_1}}.
\]
 To make sure that the Hankel matrix is of rank $\dx$, it is sufficient to select $k_1,p-1-k_1\ge \dx$.
In the setting where we know the true Markov parameters a simple Singular Value Decomposition (SVD) suffices to recover the observability and controllability matrices up to a similarity transformation. In particular letting the singular decomposition be written as
\[\Hk_{k_1,p}=\begin{bmatrix}U_1&U_2 \end{bmatrix}\begin{bmatrix}\Sigma_1&0\\0&0 \end{bmatrix}\begin{bmatrix}V^\top_1\\V^\top_2 \end{bmatrix},\]
we can select a \emph{balanced realization} $\obs_{k_1}=U_1\Sigma_1^{1/2}$, $\contr_{p-1-k_1}=\Sigma_1^{1/2}V_1^\top$. Then, from the observability/controllability matrices it is easy to recover $(\Cs,\Acls,\Hs)$ up to a similarity transformation---see for example~\cite{oymak2021revisiting}. 

However, in practice we only have access to noisy Markov parameter estimates $\hat{G}_{p,N}$, obtained for example via the least-squares identification step described above.  In this case, the corresponding Hankel matrix $\hat{\Hk}_{k_1,p}$ will also be noisy, and in particular will no longer have rank $\dx$---instead it will in general have a higher rank. In this case, a low-rank approximation step is crucial for recovering the correct observability and controllability matrices. Assume that we know the true order $\dx$ of the system. Then, we can perform SVD truncation, i.e., choose the singular vectors corresponding to the $\dx$ largest singular values. If the SVD of the noisy Hankel matrix is
\[\hat{\Hk}_{k_1,p}=\begin{bmatrix}\hat U_1&\hat U_2 \end{bmatrix}\begin{bmatrix}\hat\Sigma_1&0\\0&\hat \Sigma_2 \end{bmatrix}\begin{bmatrix}\hat V^\top_1\\\hat V^\top_2 \end{bmatrix},\]
then one solution is to keep the $\dx$-largest singular values, i.e., select $\hat\obs_{k_1,T}=\hat U_1\hat\Sigma_1^{1/2}$, $\hat\contr_{p-1-k_1,T}=\hat\Sigma_1^{1/2}\hat V_1^\top$. 

To capture the error between the true and estimated observability/controllability matrices we appeal to SVD perturbation results---more details can be found in~\cite{wedin1972perturbation}, see also \cite[Theorem 5.14]{tu2016low}. 
Essentially these results state that, for some similarity transformation $T$, the error $\opnorm{\obs_{k_1}-\hat{\obs}_{k_1,T}}$ (similarly for the controllability matrix) scales with the Markov parameter error $\opnorm{G_p-\hat{G}_{p,T}}$ as long as a robustness condition is satisfied. Ignoring dependencies on $k_1$, $p$, the robustness condition is typically of the form
\begin{equation}\label{ID_eq:SVD_robustness}
\opnorm{G_p-\hat{G}_{p,T}}\le O(\sigma_{\dx}(\Hk_{k_1,p})),
\end{equation}
namely, the Markov parameter estimation error should be smaller than the smallest singular value of the true Hankel matrix $\Hk_{k_1,p}$. 
Such a condition is a fundamental limitation of the
SVD procedure; it guarantees that the singular vectors related
to small singular values of $\Hk_{k_1,p}$ are separated from the singular
vectors coming from the noise which can be arbitrary.
While in the asymptotic regime such a condition is satisfied asymptotically, in the finite sample regime, it imposes a high sample complexity as the smallest singular value of the Hankel matrix can be very small in practice. It is an interesting open problem to look at different realization approaches or model reduction techniques so that we avoid this restrictive robustness condition. 

\begin{open_prob}[Comparison of subspace algorithms]Most results in the finite sample regime analyze the performance of the Ho-Kalman method (or similar variants)~\cite{oymak2021revisiting,tsiamis2019finite,sarkar2021finite,lee2019non}. However, in the subspace identification literature this realization approach is rarely used. Popular subspace identification algorithms, e.g., MOESP~\cite{verhaegen2007filtering} and N4SID~\cite{van2012subspace}, \rev{pre-multiply and/or post-multiply the Hankel matrix with appropriate weighting matrices, before performing the SVD step--see, for example, Section 3 in~\cite{qin2006overview}. Several asymptotic properties of such algorithmic variations have been studied before~\cite{bauer2005asymptotic}.} \rev{ However, it is an open problem to compare such algorithms using finite-sample methods. } \rev{ In particular, under finite samples, a robustness condition like~\eqref{ID_eq:SVD_robustness} should be satisfied for the SVD step to be well-behaved. Different methods lead to different robustness conditions, affecting finite-sample performance. Such robustness conditions did not appear before in asymptotic analyses, e.g. see~\cite{bauer2000impact}, since as the number of samples goes to infinity, the SVD error decays continuously.}
\end{open_prob}

\subsubsection{Overview and Limitations}
An overview of prior work can be found in Table~\ref{ID_TAB:Sys_ID_Partially}. Up to now, we studied identification of Markov parameters of both the deterministic part, i.e., $(\Cs,\Acls,\Hs)$, and the stochastic part of the system, i.e., $(\Cs,\As,\Lk)$. 
Prior work has also studied identification of exclusively the deterministic part~\cite{oymak2021revisiting,sarkar2021finite,lee2022improved,fattahi2021learning,sun2022system,djehiche2022efficient}, i.e., the Markov parameters of $(\Cs,\As,\Bs)$, where only past inputs are used as regressors. By using only inputs, these results only hold for stable systems $\rho(\As)< 1$ unless we use multiple trajectories~\cite{zheng2020non}.  In~\cite{simchowitz2019semi} it is shown that identification of non-explosive systems $\rho(\As)=1$
is possible if we also use past outputs as regressors and include a pre-filtering step in the system identification algorithm, i.e. learn an Auto-Regressive (AR) filter first before estimating the Markov parameters.
Identification of the stochastic part, i.e., the Markov parameters of $(\Cs,\As,\Lk)$, is investigated in~\cite{tsiamis2019finite}. A non-parametric approach was considered in~\cite{goldenshluger1998nonparametric}.

\paragraph{The excitation policy}
Most of the aforementioned works rely on white-noise open-loop excitation to achieve parameter recovery. Closed-loop identification under finite samples has been analyzed in~\cite{lee2019non,lale2021adaptive}, where the closed-loop controller is a linear  dynamic feedback law, potentially driven by white-noise~\cite{lee2019non}. \rev{The problem of experiment design, i.e. finding good excitation policies in the finite sample regime, remains quite open. Still, it was studied in the classical system identification literature using asymptotic tools~\cite{Ljung1999system}.}

\paragraph{The noise model}

In the case of non-Gaussian noise, the system~\eqref{ID_eq:system} and its Kalman form~\eqref{ID_eq:KF_form} have similar second moments. However, they are no longer statistically equivalent and the innovation process is no longer i.i.d. Gaussian. For this reason, some of the techniques presented above might not be applicable.  We also point out that in the case of i.i.d. sub-Gaussian noise, the results of~\cite{sarkar2021finite,simchowitz2019semi,fattahi2021learning} still hold, but only recover the deterministic part of the system. 

\paragraph{System order} The realization procedure that we presented previously requires the order of the system $\dx$  to be known. Identification of systems under unknown model order is studied in~\cite{sarkar2021finite,sun2021learning,djehiche2022efficient}.

\paragraph{Lower bounds}
\rev{Lower bounds have been studied before in the classical literature~\cite[Ch. 7]{Ljung1999system}. In the case of known system order, we can characterize the best possible parameter estimation variance among all estimators by invoking the Cramér-Rao inequality~\cite{gill1995applications}, a variant of~\nameref{sidebar-vtineq} which is studied below. One difference with Birgé's inequality is that the Cramér-Rao inequality characterizes the expected error (variance) while Birgé's inequality characterizes tail probabilities providing information about the confidence level $\delta$. Unlike fully-observed systems, existing lower bounds for partially observed systems in state-space form do not have transparent expressions in terms of system theoretic properties like the system dimension, controllability gramians, etc. This is mainly due to the non-uniqueness of state-space representations and the non-linearity of the input-to-output map with respect to the state-space parameters. }

\paragraph{Open problems in the partially observed setting}
\rev{Under the assumption that the model order is known and under certain conditions on the inputs, asymptotic optimality of several algorithms has been established. In particular, it has been shown that the prediction error method is equivalent to the maximum likelihood method~\cite[Ch. 9]{Ljung1999system}, while some subspace identification algorithms asymptotically match the maximum likelihood method under white noise excitation~\cite{bauer2005asymptotic,bauer2005comparing}. Obtaining a finite sample analog is an open problem.}

\begin{open_prob}[Optimal Sample Complexity]
\rev{
What is the optimal sample complexity in the case of partial-observability? In the case of known system order, can we match
the asymptotic performance of match maximum likelihood by a non-asymptotic analysis? What if the order is unknown? How do system theoretic parameters affect complexity?}
\end{open_prob} 

An open question is whether the optimal sample complexity should depend on the past horizon $p$. As discussed in the~\nameref{ID_sec:Regression}, this might not be the case since the number of unknowns in~$\thetas$ is independent of the horizon $p$. Some progress in this regard has already been made: in~\cite{lee2022improved}, it is shown that in the absence of process noise the sample complexity depends only logarithmically on the past horizon $p$, while retaining the $1/\varepsilon^2$ complexity rate. This is achieved by de-noising Hankel matrices at different scales. In the case of process noise, the complexity bound in~\cite{lee2022improved} still scales linearly with $p$. In~\cite{fattahi2021learning}, the sample complexity is shown to be logarithmic with $p$, at the expense of a worse $1/\varepsilon^4$ complexity rate. This is achieved by adding an $\ell_1$ regularization penalty on $G_p$ in the regression step.

To conclude, another open problem is identification of open-loop (explosively) unstable systems $(\rho(\As)>1)$ in the case of single trajectory data. While this problem is resolved in the case of fully-observed systems (under certain regularity conditions) it is still open in the case of partial observability.
\begin{open_prob}
Existing results for partially observable systems rely on stability $\rho(A_\star)\leq 1$. What, if any, are the necessary conditions for conducting open-loop unstable identification based on single trajectory of data?
\end{open_prob} 
\rev{One of the main technical difficulties in the case of unstable systems is dealing with the bias term in~\eqref{ID_eq:PO_main_regression}. If the state is increasing exponentially fast with time $k$, the bias term might not decay fast enough with $p$. In the case of non-explosive systems, two-step procedures, e.g. performing a pre-filtering step~\cite{simchowitz2019semi} or estimating components of the marginally stable subspace first~\cite{bauer2002estimating}, guarantee learnability. It is an open question whether a two-step procedure would work for (explosively) unstable systems.} 
\begin{table*}
\caption{System Identification of Partially-Observed Systems.\label{ID_TAB:Sys_ID_Partially}}
\center{
\begin{tabular}{@{}|c|c|c|c|c|c|c|c|@{}}
\hline
Paper&Trajectory&Stability&System Part&Order $\dx$&Actuation&noise\\
\hline
\cite{oymak2021revisiting,lee2022improved} & single & $\rho(\As)<1$& deterministic& known &open-loop&Gaussian 
\\
\hline
\cite{simchowitz2019semi} & single & $\rho(\As)\le 1$& deterministic& known &open-loop&sub-Gaussian
\\
\hline
\cite{tsiamis2019finite} & single & $\rho(\As)\le 1$& stochastic& known&- & Gaussian \\
\hline
\cite{sarkar2021finite} & single & $\rho(\As)<1$& deterministic& unknown &open-loop&sub-Gaussian
\\
\hline
\cite{djehiche2022efficient} & single & $\rho(\As)<1$& deterministic& unknown &open-loop&Gaussian
\\
\hline
\cite{fattahi2021learning} & single & $\rho(\As)<1$& deterministic& known &open-loop&sub-Gaussian 
\\
\hline
\cite{lee2019non,lale2021adaptive}& single & closed-loop &both& known &closed-loop&Gaussian 
\\
\hline
\cite{zheng2020non}& multiple & any& deterministic& known &open-loop&Gaussian
\\
\hline
\cite{sun2022system}& multiple & any & deterministic& unknown &open-loop&Gaussian
\\
\hline
\end{tabular}
}
\end{table*}

\section{offline control}
\label{sec:offlinecontrol}
In the previous section, we studied system identification of unknown systems under a finite number of samples. Although system identification is a problem of independent interest, our ultimate goal is to control the underlying unknown system. In this section, we connect the previous results with controlling unknown systems in a model-based framework. We also review some model-free methods. We focus on offline learning architectures, where we design the controller once after collecting the data.

This setup is very similar to the setting of episodic Reinforcement Learning (RL).
Reinforcement learning has seen tremendous success \cite{silver2016mastering,vinyals2019grandmaster}. However, most existing analyses focus on finite state and input (action) spaces. As learning methods are becoming increasingly ubiquitous even for complex continuous control tasks \cite{lillicrap2015continuous}, the gap between theory and practice has become considerable. The linear quadratic regulator (LQR) and the linear quadratic Gaussian  (LQG) problems offer a theoretically tractable path forward to reason about RL for continuous control tasks. By leveraging the theoretically tractable natures of LQR and LQG we obtain baselines and are able to quantify the performance of learning algorithms in terms of natural control-theoretic parameters. Perhaps most importantly, given the safety-critical nature of many applications  \cite{garcia2015comprehensive}, we are able to quantify what makes learning hard and when it necessarily fails. 

To make this concrete, suppose a learner (control engineer) knows that the system has dynamics of the form:
\begin{equation}
\label{eq:dynsys}
    \begin{aligned}
    x_{t+1}&=\As x_t+\Bs u_t+w_t
    \end{aligned}
\end{equation}
where, as in the previous section, we let $x_t,w_t\in\mathbb{R}^{\dx}$ be the state and process noise respectively, $u_t \in \mathbb{R}^{\du}$ be the control input. 
The dynamics matrices are $A_\star \in \mathbb{R}^{\dx\times \dx},$ and $B_\star \in \mathbb{R}^{\dx \times \du}$. In the learning task, the parameters $(A_\star,B_\star)$ are unknown to the learner. All that is known is that $(A_\star,B_\star) \in \Theta$ where $\Theta$ is some subset of parameters -- typically those corresponding to stabilizable systems.
In the offline setting, the learner is given access to $\Ntraj$ sampled trajectories of length $T$ (total of $\Ntot=\Ntraj T$ samples) from the system (\ref{eq:dynsys}) and is tasked to output a policy $\pi$ that renders the following cost as small as possible:
\begin{equation}
\label{eq:offlinecostk}
    \bar V(\theta;K) \triangleq \limsup_{T\to\infty}  \E_\theta^K \left[ \frac{1}{T}\sum_{t=0}^{T-1} \left( x_t^\top Q x_t + u_t^\top R u_t\right)\right].
\end{equation}
where expectation $\E_\theta^K$ is taken with respect to dynamics $\theta=(A,B)$ under the feedback law $u_t=Kx_t$.  In this case, it is of course known that the optimal controller is a constant state-feedback law of the form $u_t = K(A_\star,B_\star) x_t=K_\star x_t$, where the controller gain $K(A,B)$ is specified in terms of the solution $P=P(A,B)$ to a DARE: 
\begin{align}
\label{eq:dareP}
    P&=Q+A^\top P A -A^\top P B (B^\top P B + R)^{-1} B^\top P A, \\
\label{eq:dareK}
    K &=-(B^\top P B + R)^{-1} B^\top P A.
\end{align}

\subsection{Model-Based Methods} 
A classical approach to designing the optimal LQR controller for an unknown system~\eqref{eq:dynsys}, which we will revisit from a finite data perspective, is to perform system identification followed by a control design step. In RL terminology this approach is referred to as a model-based approach because we explicitly parameterize and learn the transition dynamics, which are then used compute a policy. In particular, suppose that we have obtained estimates $(\widehat A,\widehat B)$ of $\theta_\star = (A_\star,B_\star)$ and that these estimates are guaranteed to be $\e$-accurate, i.e., $\max \{ \opnorm{A_\star-\widehat A},\opnorm{B_\star-\widehat B}\}\leq \e$. Such estimates can be acquired and guaranteed to satisfy the desired accuracy level (with high probability) by leveraging the results of  the above discussion on \nameref{sec:samplecomplexityub}. Based on the system estimates, we can either apply certainty equivalent control or design a robust controller using the error information $\varepsilon$.

\subsubsection{Certainty Equivalence}
The certainty equivalent (CE) approach is to simply use the estimates $(\widehat A,\widehat B)$ as if they were the ground truth and play the controller $\widehat K = K(\widehat A,\widehat B)$.

The situation described above is the precisely that analyzed in \citet[Theorem 2]{mania2019certainty}. They demonstrate that the controller $\widehat K = K(\widehat A,\widehat B)$ enjoys the sub-optimality guarantee
\begin{align}
\label{eq:maniace}
    \bar V(\theta_\star;\widehat K)- \bar V(\theta_\star; K_\star) \leq \textsf{poly}_{\theta_\star} \e^2
\end{align}
where $\textsf{poly}_{\theta_\star}$ denotes a quantity polynomial in system quantities such as $\opnorm{P_\star}$ and the spectral radius of the optimal closed-loop dynamics $A_\star+B_\star K_\star$---one can view the term $\textsf{poly}_{\theta_\star}$ as capturing that systems with well-conditioned closed-loop behavior are easier to learn to control. Similar guarantees can also be provided for the partially observed LQG setting in which the entire linear dynamic controller is estimated from data \cite[Theorem 3]{mania2019certainty}.

It is important to recognize, however, that guarantee~\eqref{eq:maniace} comes with the caveat that the accuracy $\e$ needs to be small enough so that the controller $\widehat K$ can be shown to be stabilizing for the instance  $\theta_\star=(A_\star,B_\star)$. \citet{mania2019certainty} provide sufficient conditions on the accuracy $\e$ in terms system parameters by leveraging \nameref{sidebar-riccati}. The dependence on $\e$ in inequality \eqref{eq:maniace} is optimal and in fact it can be shown that \emph{for almost every} experiment consisting of input-state data $\{(x_0,u_0),\dots,(u_{\Ntot-1},x_{\Ntot})\}$, the least squares estimator described above in combination with certainty equivalent control is optimal \citep[Theorem 2.1]{wagenmaker2021task} in that up to universal constants there exists no better strategy. In fact, we will later see that the CE approach is also the best known strategy in the more challenging \nameref{sec:regmin} setting.

Combining guarantee~\eqref{eq:maniace} with the~\nameref{sec:samplecomplexityub} of the previous section, we can obtain end-to-end guarantees for the offline learning of the optimal LQR controller. In particular, we obtain that the suboptimality gap decreases at least as fast as $\tilde{O}(1/\Ntot)$. However, as stated earlier, this result assumes that the number of samples is large enough such that the CE controller $\hat{K}$ is stabilizing for the original system, which may require a large burn-in time.

\begin{sidebar}{Riccati Equation Perturbation Theory}
\section[Riccati Equation Perturbation Theory]{}\phantomsection
   \label{sidebar-riccati}
\setcounter{sequation}{0}
\renewcommand{\thesequation}{S\arabic{sequation}}
\setcounter{stable}{0}
\renewcommand{\thestable}{S\arabic{stable}}
\setcounter{sfigure}{0}
\renewcommand{\thesfigure}{S\arabic{sfigure}}

\sdbarinitial{T}o provide a guarantee of the form \eqref{eq:maniace}  for the CE approach we need to guarantee that small errors in the estimates $\max \{ \opnorm{A_\star-\widehat A},\opnorm{B_\star-\widehat B}\}\leq \e$ translate to small errors in Riccati equation quantities \eqref{eq:dareP}-\eqref{eq:dareK}. Key to achieving such guarantees is an operator-theoretic proof strategy due to \cite{konstantinov1993perturbation}. Roughly, the idea is  to construct a map $\Phi$ of which the error $P_\star -\widehat P$ is the unique fixed point over a set of elements with small norm. A more detailed account can be found in \cite[Section 4.1]{mania2019certainty}. We also note that \cite[Section 3]{simchowitz2020naive} has recently developed an alternative ODE approach which gives tighter bounds in terms of system-theoretic parameters.

\end{sidebar}

\subsubsection{Robust Control Methods}
While the CE controller is optimal when the model error $\varepsilon$ is very small, there are nevertheless many cases of interest where only a coarse model is available and the model error is too large to guarantee that the CE controller is stabilizing~\cite{dean2020_on_the_sample}. In such settings, an alternative is to design a robust controller which stabilizes all possible systems consistent with the model estimates and error bounds. In~\cite{tu2017non}, the problem of robust control from coarse system identification was studied in the non-asymptotic regime. In~\cite{dean2020_on_the_sample} a robust control scheme based on System Level Synthesis (SLS)~\cite{ANDERSON2019364} is introduced which uses finite sample model error information. 
The aforementioned robust control designs are safer than the CE controller in general. However, the cost of this robustness is that the resulting controller suboptimality guarantees are worse. Contrary to~\eqref{eq:maniace}, they enjoy suboptimality guarantees of the order of 
\begin{align}
\label{OFF_eq:robust_suboptimal}
    \bar V(\theta_\star;\widehat K)- \bar V(\theta_\star; K_\star) \leq \textsf{poly}_{\theta_\star} \e,
\end{align}
where $\widehat{K}$ is the robust controller. It is unknown whether this suboptimality is inherent or an artefact of the analysis.
\rev{SLS controllers can also be deployed in the case of state/input constraints~\cite{dean2019safely} as well as partially observed systems~\cite{boczar2018finite}. An alternative Input-Output Parameterization (IOP) framework was adapted in~\cite{furieri2022near} to deal with uncertain partially observed systems. }

\subsection{Model-Free Methods}
Model-free methods, in which (essentially) no structural information about the problem is used to derive a learning-based policy, are very popular in the RL literature. The most basic class of such methods are policy gradient methods, which we discuss next in the context of the LQR problem.

\subsubsection{Policy Gradient Methods}
Policy gradient methods work exactly as their name advertises: they run (stochastic) gradient descent on a controller-parameterization with respect to the cost \eqref{eq:offlinecostk}. To make this concrete, let us for simplicity first discuss the state-feedback setting in which $C_\star = I_{\dx}$ and $v_t=0$. In light of the form \eqref{eq:dareP}-\eqref{eq:dareK} of the optimal policy, it appears reasonable to parametrize the cost \eqref{eq:offlinecostk} by linear controllers of the form $u_t= Kx_t$ and run our descent steps on matrices $K\in \mathbb{R}^{\du\times \dx}$. 

\paragraph{Do Exact Gradients Converge?}
Assume for the moment that we have oracle access to \emph{exact gradients} and that we are able to run (non-stochastic) gradient descent on the cost function \eqref{eq:offlinecostk}: $$K_{j+1} = K_j- \nabla_K V_T(\theta;K)\Big|_{K=K_j}.$$ It is not obvious that such an algorithm will work, as even in this simplified setting, there are two potential obstacles to convergence: 1) the cost function \eqref{eq:offlinecostk} is non-convex in $K$; and 2) the cost function \eqref{eq:offlinecostk} is not globally smooth---in fact, it is not even finite for those $K$ that do not stabilize the system \eqref{eq:dynsys}. Thankfully, the LQR objective \eqref{eq:offlinecostk} satisfies \nameref{sidebar-opt-pl-smoothness} which are entirely sufficient. These weaker conditions were first established by \citet{fazel2018global} who showed that if initialized with a stabilizing controller $K_0$, after only $O(\log 1/\e)$ iterations, (non-stochastic) gradient descent outputs a controller $\tilde K$ satisfying \begin{equation}
\label{eq:glinconv}
    \bar V(\theta,\tilde K)-\min_K \bar V(\theta,K) \leq \e.
\end{equation}
It should be noted that \cite{fazel2018global} consider a slightly different cost function than the cost considered here. Namely, they consider the infinite horizon case with $w_t=0$ and only the initial condition $x_0$ is allowed to be random. However, the infinite horizon and ergodic average cost functions are almost identical (as functions of $K$), and it is straightforward to verify that the convergence guarantee mentioned above remains true with only minor modifications to problem-specific constants when applied to the ergodic average cost \eqref{eq:offlinecostk}. Having established that the exact gradient method converges, \citet{fazel2018global} also showed that a method based on zero-order gradient estimates also converges. However, their results only apply to the noiseless setting with random initial condition. By contrast, \cite{hambly2021policy} analyze a noisy finite horizon setting and show that such methods still provably converge. We also point out that the assumption of an initial stabilizing controller mentioned above can be removed with a more sophisticated gradient strategy \cite{perdomo2021stabilizing}. We refer the reader to the recent survey \cite{hu2022towards} for a more comprehensive overview of policy gradient methods.

\subsubsection{Fundamental Limits and Model-Based versus Model-Free}
Given the optimality of the CE controller in the offline LQR setting, it is natural to wonder whether similar guarantees are achievable by model-free methods based on policy gradients. To this end \citet{tu2019gap} study a simplified version of LQR \eqref{eq:offlinecostk} in which $R=0$ and the optimal solution is of the form $K_\star =- B_\star^\dagger A_\star$. In this simplified scenario they compute asymptotically exact expressions for the risk of CE and a stochastic policy gradient method (REINFORCE), and show that that there is a polynomial gap in the problem dimension in their respective sample complexities, with CE outperforming REINFORCE. The fundamental limits of policy gradient methods are further investigated and related to various system-theoretic quantities in \cite{ziemann2022policy}.

\begin{sidebar}{LQR, Polyak-Łojasiewicz and Approximate Smoothness}
\section["weaker versions" of convexity and smoothness]{} \phantomsection
   \label{sidebar-opt-pl-smoothness}
\setcounter{sequation}{0}
\renewcommand{\thesequation}{S\arabic{sequation}}
\setcounter{stable}{0}
\renewcommand{\thestable}{S\arabic{stable}}
\setcounter{sfigure}{0}
\renewcommand{\thesfigure}{S\arabic{sfigure}}

\sdbarinitial{W}hile the LQR objective is not convex, the objective \eqref{eq:offlinecostk} satisfies the so-called Polyak-Łojasiewicz (PL) condition. Namely, \citet[Lemma 3]{fazel2018global} show that as long as the tuple $(A,\sqrt{\Sigma_W})$ is controllable, the following PL condition holds:
\begin{sequation}
\label{eq:plcond}
    \bar V(\theta, K)-\min_K \bar V(\theta,K) \leq \lambda \| \nabla_K\bar V(\theta, K)\|_F^2
\end{sequation}for some problem-specific constant $\lambda>0$. PL Conditions such as inequality \eqref{eq:plcond} are known to be sufficient alternatives to (strong) convexity in the optimization literature \citep{polyak1963gradient,karimi2016linear}. In particular, condition \eqref{eq:plcond} enforces that any stationary point is a global minimizer, as is the case for convex functions. An alternative perspective on the condition \eqref{eq:plcond} is offered in \cite{sun2021learning}, in which it is shown to be a consequence of the existence of a convex re-parametrization for the LQR objective. 

Similarly, even though the objective \eqref{eq:offlinecostk} is not globally smooth, it is sufficiently regular in that:
\begin{align*}
    \bar V(\theta, K)- \bar V(\theta,K_\star) = \langle \nabla_K\bar V(\theta, K), K-K_\star\rangle_F + O(\| K-K_\star\|_F^2)
\end{align*}
in a neighborhood of the optimal policy $K_\star$.

In combination, these properties can be used to verify that if gradient descent is initialized with a stabilizing controller, its updates remain stable and converge to the global optimum at the rate \eqref{eq:glinconv}.
\end{sidebar}

\section{online control}
\label{sec:regmin}

Having discussed episodic RL tasks through the lens of control, we now turn our attention to the more technically challenging setting of online adaptive control.  We will rely on the notion of \emph{regret} to quantify the performance of an online algorithm.

Just as in \nameref{sec:offlinecontrol}, we suppose the system has dynamics are of the form:
\begin{equation}
\label{eq:dynsysac}
    \begin{aligned}
    x_{t+1}&=A_\star x_t+B_\star u_t+w_t\\
    y_t&=C_\star x_t+v_t
    \end{aligned}
\end{equation}
where $x_t,w_t\in\mathbb{R}^{\dx}, u_t \in \mathbb{R}^{\du}, y_t,v_t\in \mathbb{R}^{\dy}$ and $A_\star \in \mathbb{R}^{\dx\times \dx}, B_\star \in \mathbb{R}^{\dx \times \du}$ and $C_\star \in \mathbb{R}^{\dy\times \dx}$. However, in contrast to the \nameref{sec:offlinecontrol} setting, the learner now interacts iteratively with only a single trajectory ($\Ntraj=1$, $T=\Ntot$) from the system (\ref{eq:dynsysac}). The parameters of $(A_\star,B_\star,C_\star)$ are as before unknown to the learner.

For simplicity, we will assume that $\{w_t\}$ and $\{v_t\}$ are mutually independent i.i.d. sequences of mean zero sub-Gaussian random variables, with covariance matrices $\Sigma_w$ and $\Sigma_v$ respectively. Most of the current literature focuses on the LQR setting, where we take $C=I_{\dx}$ and $v_t=0$. Relatively less is known about regret minimization for the partially observed setting (in which case the noise sequences are Gaussian).

In either setting, the goal in the adaptive LQR and LQG problems is to regulate the system (\ref{eq:dynsysac}) using a policy $\pi$ so as to render the following cost functional as small as possible:
\begin{equation}
\label{eq:lqrcost}
    V_T^\pi(\theta) \triangleq \E_\theta^\pi \left[x_T^\top Q_T x_T + \sum_{t=0}^{T-1} x_t^\top Q x_t + u_t^\top R u_t  \right]
\end{equation}
where $\E_\theta^\pi$ stands for expectation with respect to dynamics $\theta=(A,B,C)$ under policy $\pi$ and where $(Q,Q_T,R)$ are positive definite weighting matrices. The difficulty of the task arises from the fact that the parameter $\theta$ is assumed a priori unknown, and hence the optimal cost $V_T^\star(\theta_\star) \triangleq \inf_{\pi \in }V_T^\pi(\theta)$ can not be realized. Instead, one seeks to design a policy (algorithm) $\pi$ with small regret. 

\begin{tcolorbox}[
 colframe=red!70!white,
 colback=red!9!white,
 arc=8pt,
 breakable,
 left=1pt,right=1pt,top=1pt,bottom=1pt,
 boxrule=0.3pt,
 ]
\textbf{Regret.} The regret of an algorithm measures the cumulative suboptimality accrued over the entire time horizon as compared to the optimal policy:
\begin{equation}
\label{eq:regret}
    \Regret^\pi_T(\theta) \triangleq x_T^\top Q_T x_T + \sum_{t=0}^{T-1} x_t^\top Q x_t + u_t^\top R u_t - V_T^\star(\theta)
\end{equation}
where the law of $\{x_t,u_t\}_{t=0}^{T}$ is specified by $(\theta,\pi)$. Alternatively, one may be interested in the expected regret:
\begin{equation}
\label{eq:expectedregret}
    \E \Regret^\pi_T(\theta) = V_T^\pi(\theta) -V_T^\star(\theta).
\end{equation}
 \end{tcolorbox}

We note that the regret is a random quantity whereas the expected regret is not---however, in either case the interpretation is that one seeks to design a policy which has small cumulative suboptimality as compared to the optimal policy $\pi_\star(x)=K_\star x$, which can be computed via Riccati equations \eqref{eq:darrP}-\eqref{eq:darrK}.  \rev{Abstracting slightly, the regret of an algorithm can be thought of as the rate of convergence of an adaptive algorithm (cf. 
\eqref{eq:expectedregretrep2}). Moreover, it quantifies the dual nature of control \citep{feldbaum1960dual1,feldbaum1960dual2} (in RL terminology: the exploration-exploitation trade-off). We will see in the sequel that for an algorithm to have low regret it necessarily must generate a sufficiently rich experiment. At a high level, by relating \eqref{eq:regret} (or \eqref{eq:expectedregret}) to quantities of interest such as the time horizon $T$, dimensional factors and system-theoretic quantities we gain understanding of the statistical properties of adaptation and under which circumstances adaptation---if only in an idealized environment---is easy or hard. We should also point out that in the formulation \eqref{eq:regret}-\eqref{eq:expectedregret} we compete with a policy that has good average case performance (LQR) but does not necessarily take into robust or stability margins. While certainly important, in this survey we do not cover robustness aspects of adaptive methods but rather emphasize their statistical analysis.}

\paragraph{State Feedback Systems}
For state feedback systems ($C_\star =I_{\dx}, v_t =0$), it has been shown by \citet{simchowitz2020naive} that \nameref{sec:certaintyequivalence} with naive exploration (additive Gaussian noise injected into the control input) attains with probability $1-\delta$:
\begin{align*}
    R_T^\pi(\theta_\star) \leq c_{\mathsf{sys}} \sqrt{\dx\du^2 T \log (1/\delta)}
\end{align*}
for a system-dependendent constant $c_{\mathsf{sys}}>0$ and provided that $T$ is sufficiently large (polynomial in dimension and system-dependent quantities).

Their result refined an earlier result of \cite{mania2019certainty} and essentially settled the question of what the optimal dependence on system dimensions and time horizon is. A recent result due to \citet{jedra2022minimal} also shows that, up to logarithmic factors,  the same rate can be attained in expectation $\E R_T^\pi(\theta_\star) =\tilde O\left(\sqrt{\dx\du^2 T}\right)$. \citet{simchowitz2020naive} also provide a matching lower bound with $\sup_{\theta \in B(\theta_\star,\e)} \E R_T^\pi(\theta) = \Omega \left(\sqrt{\dx\du^2 T}\right) $. However, characterizing the optimal dependence on the system parameters $(A_\star, B_\star)$ is still open. There is for instance polynomial gap in the best known upper bounds \cite{simchowitz2020naive} the best known lower bounds \cite{ziemann2022regret} in regards to the dependence on $P_\star=P(A_\star,B_\star)$ (recall \eqref{eq:dareP}). A summary of the state of the art for both state feedback and partially observed systems is given in \autoref{table:adaptsum}.

\begin{table*}
\center
\caption{Summary of Results: Regret Minimization in Adaptive Control -- state of the art in blue.}
\label{table:adaptsum}
\begin{tabular}{|c |c|c|c|c| } 
 \hline
 Paper & Setting & Method & Upper Bound & Lower Bound \\ 
 \hline
\cite{abbasi2011regret}  & SF: $(A,B)$ unknown & Optimism &$ \tilde O(\sqrt{T})  $ but intractable&   \\ 
\hline
\cite{dean2018regret}  &SF: $(A,B)$ unknown & CE& $\tilde O(T^{2/3})$ &  \\ 
 \hline 
 \cite{faradonbeh2020input} & & CE &  &\\
 \cite{mania2019certainty} &SF: $(A,B)$ unknown & CE  & $ \tilde O(\sqrt{T})$ &  \\ 
 \cite{cohen2019learning} &  &Optimism& &  \\ 
 \hline
 \cite{simchowitz2020naive} &SF: $(A,B)$ unknown & CE & ${\color{blue}O( \sqrt{\dx \du^2 T})}$ & $ \Omega( \sqrt{\dx \du^2 T})$ \\ 
 \hline
\cite{cassel2020logarithmic}  &SF: $A$ unknown & CE& $\color{blue} \tilde O(\log T)$ & \\ 
& $b$ scalar unknown & CE& & $\Omega(\sqrt{T})$\\
\hline
 \cite{simchowitz2020improper}&PO: $(A,B,C)$ unknown &Gradient& $\color{blue}\tilde O(\sqrt{T})$ &  \\ 
 \hline
  \cite{ziemann2022regret} &SF: $(A,B)$ unknown & & & $\color{blue}\Omega(\sqrt{\dx \du^2 T})$  \\

   &PO: $(A,B,C)$ unknown & & & $\color{blue}\Omega(\sqrt{T})$  \\
 \hline
   \cite{Tsiamis2022Learning} &SF: $(A,B)$ unknown & CE & $\color{blue} O(\exp(\kappa)\sqrt{\dx\du^2 T}) $ & $\color{blue}\Omega\left(\sqrt{\frac{1}{\dx}2^{\kappa} T}\right)$  \\ 
 \hline
\end{tabular}
\end{table*}

\subsection{Certainty Equivalence}
\label{sec:certaintyequivalence}
The key algorithmic idea to solve the regret minimization problem for LQR is again certainty equivalence (CE). The idea dates back to the late 50s \cite{simon1956dynamic, feldbaum1960dual1,feldbaum1960dual2} and was first analyzed in the context of adaptive control of linear models by \citet{aastrom1973self} in 1973. Initially, the emphasis was solely on asymptotic average cost optimality, corresponding to sublinear regret, $\Regret_T^\pi =o(T)$, in our formulation. Regret minimization was introduced to the adaptive control literature roughly a decade later by \citet{lai1986asymptotically}.

Online CE LQR control takes continuously updated parameter estimates $(\widehat A,\widehat B, \widehat C)$ of $(A_\star,B_\star,C_\star)$ as inputs and then solves the dynamic programming problem for these estimates as if they were the ground truth. For LQR, the dynamic programming solution has a closed form solution in terms of the (discrete algebraic) Riccati recursion \eqref{eq:darrP}-\eqref{eq:darrK} which can be solved efficiently by numerical schemes. The resulting controller is then used to regulate the system. 

To see why the CE strategy is successful in LQR we note the following elementary relation between expected regret and the Riccati recursion \citep{ziemann2022regret}:
\begin{align}
\label{eq:expectedregretrep}
    \E R_T^\pi(\theta) = \sum_{t=0}^{T-1} \E_{\theta}^\pi \left[ (u_t - K_tx_t)^\top (B^\top P_{t+1} B+R)(u_t - K_tx_t)\right]
\end{align}
where $\theta=(A,B)$, $P_t=P_t(\theta)$  and $K_t=K_t(\theta)$  are given by
\begin{align}
\label{eq:darrP}
    P_{t-1} &=Q+A^\top P_t A -A^\top P_t B (B^\top P_t B + R)^{-1} B^\top P_t A, \\
\label{eq:darrK}
    K_t &=-(B^\top P_t B + R)^{-1} B^\top P_t A,
\end{align}
and where the terminal condition is $P_T=Q_T$. We further denote the steady state versions of the recursion (\ref{eq:darrP})-(\ref{eq:darrK}) by $P(A,B)$ and $K(A,B)$. It will be convenient to denote $P_\star \triangleq P(A_\star,B_\star)$ and $K_\star \triangleq K(A_\star,B_\star)$.

Equation (\ref{eq:expectedregretrep}) follows from the "completing-the-square" proof of LQR optimality, cf. \cite[Theorem 11.2]{soderstrom2002discrete}. Crucially, for naive exploration policies of the form $\pi: u_t = \widehat K_t x_t+\eta_t$, with $\{\eta_t\}$ a mean zero sequence of exploratory noise, independent of all other randomness, equation (\ref{eq:expectedregretrep}) becomes
\begin{multline}
\label{eq:expectedregretrep2}
     \E R_T^\pi(\theta) =\E_{\theta}^\pi \sum_{t=0}^{T-1} \eta_t^\top \eta_t \\
      +\sum_{t=0}^{T-1} \E_{\theta}^\pi \left[ x_t^\top (K_t-\widehat K_t)^\top  (B^\top P_{t+1} B+R)(K_t-\widehat K_t)x_t\right].
\end{multline}
Equation (\ref{eq:expectedregretrep2}) shows that the expected regret of a CE policy is a quadratic form in the estimation error $\widehat K_t - K_t$.  Moreover, by a stability argument it suffices to use the steady-state versions of the Riccati recursion (\ref{eq:darrP}-\ref{eq:darrK}). This suggests that the CE strategy with $\widehat K_t = K(\widehat A, \widehat B)$ can be shown to be successful provided that one shows that the
\begin{enumerate}
    \item estimates $(\widehat A, \widehat B)$ are consistent estimators of the true dynamics; and
    \item map $(A,B) \mapsto K_t(A,B)$ is sufficiently smooth in the parameters $(A,B)$; and
    \item policy $\pi$ is stabilizing in that the state process $x_t$ does not become too large.
\end{enumerate}
Analogous reasoning is applicable in the high probability regret setting, but becomes a little more involved, see \cite[Lemma 5.2]{simchowitz2020naive}.

Before we proceed one remark is in order: equation \eqref{eq:expectedregretrep2} suggests that $\E \Regret_T^\pi(\theta) =O(log T)$ should be possible. Namely, we noted in the \nameref{sec:finsampsysid} that the identification errors generally decline as $O(1/\sqrt{t})$, where $t$ is the number of samples collected so far.  As the suboptimality bound~\eqref{eq:maniace} is quadratic in the identification error, the square errors decline as $O(1/t)$ and the regret induced will scale as the sum of $1/t, t=0,\dots, T-1$, which is of order $\log T$. We will soon ask: "\nameref{par:why}" and see that logarithmic regret is not possible in general for reasons of closed-loop identifiability.

\begin{sidebar}{Optimism and Thompson Sampling}
\section[Optimism and Thompson Sampling]{} \phantomsection
   \label{sidebar-opt-thompson}
\setcounter{sequation}{0}
\renewcommand{\thesequation}{S\arabic{sequation}}
\setcounter{stable}{0}
\renewcommand{\thestable}{S\arabic{stable}}
\setcounter{sfigure}{0}
\renewcommand{\thesfigure}{S\arabic{sfigure}}

\sdbarinitial{A}lternative expoloration strategies include Optimism and Thompson sampling. Indeed, the first complete treatment of regret minimization in LQR, due to \citet{abbasi2011regret}, relies on the principle of \emph{optimism in the face of uncertainty} (OFU). Just as in the CE approach discussed in the main text OFU is based on constructing parameter estimates $(\widehat A,\widehat B)$. However, OFU also maintains a (tuned) confidence interval for these estimates. The adaptive control law is then obtained by selecting the most \emph{optimistic} parameter and CE control law---those resulting in the lowest estimated cost--- in this confidence interval. The original algorithm of \cite{abbasi2011improved} was not computationally tractable, but this was later remedied by \cite{abeille2020efficient}. A related method, \emph{Thompson Sampling}, is studied in \cite{ouyang2017control, abeille2018improved}. 

We note in passing that even though these strategies in principle are more sophisticated, to date, the tightest bounds have been proven for the simple input perturbation approach described in the main text \cite{simchowitz2020naive}.
\end{sidebar}

\paragraph{Why do we need Exploration?}
\label{par:why}
In the sketch of the certainty equivalent approach presented above we mentioned that one typically requires a perturbation $\eta_t$ of the input $u_t$. The most common exploration strategy, known as $\e$-greedy exploration, uses simple additive perturbations to the control policy, yielding inputs of the form $u_t = K_t x_t + \eta_t$ as above. More intricate exploration strategies are however possible, as described in the sidebar on \nameref{sidebar-opt-thompson}. To understand why such perturbations are necessary, consider again the least-squares algorithm~\eqref{ID_eq:ls_algorithm}.  
Recall that the error of the estimator $\widehat \theta_s = (\widehat A_s,\widehat B_s)$ satisfies the following equation:
{\medmuskip=0mu\thickmuskip=1mu\thinmuskip=5mu
\begin{align}
\label{eq:LSEinAC}\arraycolsep=1.4pt
    \widehat \theta_s  -\theta_\star = \left(\sum_{t=0}^{s-1} w_{t} \begin{bmatrix} x_t^\top & u_t^\top \end{bmatrix} \right)\left( \sum_{t=0}^{s-1} \begin{bmatrix} x_t \\ u_t \end{bmatrix}\begin{bmatrix} x_t^\top & u_t^\top \end{bmatrix} \right)^{-1}
\end{align}
}provided the matrix inverse on the right hand side of equation (\ref{eq:LSEinAC}) exists. As mentioned above, as long as the covariates do not grow more than polynomially with the time horizon, it can be shown using the theory of \nameref{sidebar-self-normalized} that the rate of convergence of $\widehat \theta_s -\thetas$  is dictated by the smallest eigenvalue of the covariates matrix
{\medmuskip=0mu\thickmuskip=1mu\thinmuskip=5mu
\begin{align}
\label{eq:opnormconvlseac}\arraycolsep=1.4pt
  \opnorm{\widehat \theta_s - \theta_\star} =\tilde O  \left[ \lambda^{-1}_{\min} \left(\sqrt{\left( \sum_{t=0}^{s-1} \begin{bmatrix} x_t \\ u_t \end{bmatrix}\begin{bmatrix} x_t^\top & u_t^\top \end{bmatrix} \right)}\right)\right].
\end{align}}Suppose for the moment $u_t \approx K_\star x_t$ in equation (\ref{eq:opnormconvlseac}). In this case the matrix
\begin{align}
\label{eq:optpolsing}
     \sum_{t=0}^{s-1} \begin{bmatrix} x_t \\ u_t \end{bmatrix}\begin{bmatrix} x_t^\top & u_t^\top \end{bmatrix} \approx \sum_{t=0}^{s-1} \begin{bmatrix} I_{\dx} \\K_\star  \end{bmatrix} x_t x_t^\top  \begin{bmatrix} I_{\dx} & K_\star^\top \end{bmatrix}
\end{align}
is nearly singular. To see this, note that $\begin{bmatrix} I_{\dx} & K_\star^\top \end{bmatrix}^\top$ is a tall matrix---the outer product of tall matrices is singular. Thus, the error (\ref{eq:opnormconvlseac}) diverges if the policy is too close to the optimal policy $K_\star$, i.e., the true parameters $A_\star$ and $B_\star$ are not identifiable under the optimal closed-loop policy $K_\star$.  In fact, this lack of identifiability is true under any policy of the form $u_t = Kx_t$. 

Alternatively, the need for exploration can be seen by noting that for every perturbation $\Delta \in \mathbb{R}^{\dx\times\du}$ and $(A(\Delta),B(\Delta))$ of the form $A(\Delta) = A_\star -s\Delta K_\star, B(\Delta) = B+s\Delta$ ($s \in \mathbb{R})$ the closed loop systems $A_\star + B_\star K_\star$ and $A(\Delta)+B(\Delta)K_\star$ are identical:  $A_\star + B_\star K_\star = A(\Delta)+B(\Delta)K_\star$ for all such $\Delta,s$. As such, from observing trajectories generated by the two systems
\begin{align*}
    x_{t+1}& = (A_\star + B_\star K_\star)x_t +w_t,\\
    x_{t+1}&=(A(\Delta)+B(\Delta)K_\star)x_t+w_t
\end{align*}
it is impossible to distinguish between them. The reasoning above indicates that in order to obtain estimates that convergence sufficiently quickly to the true parameters $(A_\star, B_\star)$, exciting inputs that lead to exploration away from the optimal policy $K_\star$ are necessary.

\paragraph{Do we actually need to identify the true parameters $(A_\star, B_\star)$?}
The answer to this question is in the affirmative. To see this, we recall from
\cite[Lemma 2.1]{simchowitz2020naive} that \begin{multline}
\label{eq:simchoprop}
    \frac{d}{ds} K(A_\star -s\Delta K_\star, B_\star + s\Delta)\Big|_{s=0}\\
    = -(B_\star^\top P_\star B_\star +R)^{-1}\Delta^\top P_\star (A_\star +B_\star K_\star).
\end{multline}
As long as $(A_\star +B_\star K_\star)$ in the matrix on the right hand side of equation (\ref{eq:simchoprop}) is nonzero this implies that there exists a confusing parameter variation (which is not closed-loop distinguishable) that has a different optimal policy. Hence, one necessarily must identify the true parameters $A_\star$ and $B_\star$ in the adaptive control problem.

\begin{tcolorbox}[
 colframe=green!70!white,
 colback=green!9!white,
 arc=8pt,
 breakable,
 left=1pt,right=1pt,top=1pt,bottom=1pt,
 boxrule=0.3pt,
 ]
 \label{box:cliden}
\textbf{A historical tangent on identifiability.}
Closed-loop identifiability issues are well-known in the system identification literature \citep{lin1985will, gevers1986optimal, polderman1986necessity}. Indeed, in the LQR setting, \citet{polderman1986necessity} gives an elegant geometric argument showing that the true parameters need to be identified. It is also interesting to note that, precisely because the minimum variance controller ($Q=I,R=0$) is closed-loop identifiable \citep{lin1985will} (in contrast to the more general LQR controller), logarithmic regret can be achieved in this setting~\citep{lai1986asymptotically}. Reiterating the point above: the reason for the necessity of the "exploratory signals" $\eta_t$ in equation \eqref{eq:expectedregretrep2} is precisely a lack of closed-loop identifiability.
 \end{tcolorbox}

Returning to our estimation guarantee (\ref{eq:opnormconvlseac}), we note that an i.i.d. sequence $\eta_t$ of rescaled isotropic noise of magnitude (standard deviation) $t^{-\alpha}$ is sufficient to guarantee parameter recovery at the rate: $ \opnorm{\widehat \theta_t - \theta_\star} = \tilde O(t^{\alpha-1/2})$. In this case, smoothness (combined with a naive taylor expansion) suggests that $\opnorm{K(\widehat A_t,\widehat B_t)-K_\star } =\tilde O(t^{\alpha-1/2})$. Balancing the two terms in equation (\ref{eq:expectedregretrep2}) we see that $\alpha=1/4$ leads to $R_T = \tilde O(\sqrt{T})$, which is optimal. While the reasoning above about the necessity of the perturbations $\eta_t$ is entirely heuristic, it can be made formal and will be discussed further in the section on regret lower bounds below.

\subsection{Regret Lower Bounds}
We now argue that the scaling $R_T^\pi = \Theta (\sqrt{\dx \du^2 T})$ is optimal for state feedback systems by finding matching lower bounds. The modern approach to lower bounds, or fundamental performance limits, for sequential decision making problems seeks to characterize local minimax lower bounds. Such bounds quantify statements of the form "there exists no algorithm which uniformly outperforms a certain fundamental limit across a small (local) neighborhood of problem parameters". For the regret minimization problem such lower bounds typically take the form:
\begin{align}
\label{eq:genericlowerbound}
   \sup_{\theta \in B(\theta_\star,\e)}  \E R_T^\pi(\theta) \geq f(\theta_\star,\e,T)
\end{align}
for some $\e>0$, some function $f$ and for every (causal) policy $\pi$. The lower bound (\ref{eq:genericlowerbound}) states that the worst case expected regret over a neighborhood of the true parameter is lower bounded by some function of the instance parameter $\theta_\star$ and the horizon $T$. The appearance of $\sup_{\theta \in B(\theta_\star,\e)}$ in inequality (\ref{eq:genericlowerbound}) is not restrictive---while such lower bounds are "worst case" one can typically allow for $\e \to 0$. In other words, such lower bounds are applicable to all algorithms which are in some sense robust to infinitesimal perturbations in the model parameter $\theta_\star$, a rather mild criterion. Put yet differently, a lower bound of the form (\ref{eq:genericlowerbound}) for vanishing $\e\to 0$ states that there exists no algorithm which uniformly outperforms the lower bound in an infinitesimal neighborhood.

\begin{sidebar}{Van Trees' Inequality and Fisher Information}
\section[Van Trees' Inequality]{} \phantomsection
   \label{sidebar-vtineq}
\setcounter{sequation}{0}
\renewcommand{\thesequation}{S\arabic{sequation}}
\setcounter{stable}{0}
\renewcommand{\thestable}{S\arabic{stable}}
\setcounter{sfigure}{0}
\renewcommand{\thesfigure}{S\arabic{sfigure}}

\sdbarinitial{V}an Trees' inequality is an MMSE  lower bound for Bayesian estimation problems. Suppose the learner seeks to estimate a smooth function $\psi(\theta)$ of a parameter $\theta$. The learner is given access to a sample $Z$ drawn conditionally from a density $p(z | \theta)$ and has access to a prior $\lambda(\theta)$. To state Van Trees' inequality, define the Fisher Information as
\begin{equation*}
    \I_{p}(\theta) \triangleq \int  \left[\nabla_\theta \log p(z|\theta) \right] \left[ \nabla_\theta \log p(z|\theta) \right]^\top p(z|\theta) dz
\end{equation*}
and the prior information as:
\begin{align*}
    \J(\lambda) \triangleq \int  \left[\nabla_\theta \log \lambda(\theta) \right] \left[ \nabla_\theta \log \lambda(\theta) \right]^\top \lambda(\theta) d\theta.
\end{align*}

Under a few relatively mild regularity conditions, Van Trees' Inequality states that any estimate using $Z$ satisfies the lower bound
\begin{multline*}
    \E \left[ (\hat \psi -\psi(\theta))(\hat \psi -\psi(\theta))^\top\right] \succeq \\
    \E \nabla_\theta \psi(\theta) \left[ \E \I_p(\theta)+\J(\lambda) \right]^{-1} \E [\nabla_\theta \psi(\theta)]^\top.
\end{multline*}
where $\E$ denotes expectation with respect to $p(y,\theta) = p(y|\theta) \lambda(\theta) $.

For our purposes, it is important to note that the Fisher Information $\I_p(\theta)$ for $Z=\{x_t,u_t\}_{t=0}^{T-1}$ with $x_{t+1}=Ax_t+Bu_t +w_t$ and $\theta = \VEC (A,B)$ is equal to
\begin{align*}
    \I_p(\theta)= \E \left[\sum_{t=0}^{T-1}  \left( \begin{bmatrix} x_t \\ u_t \end{bmatrix}\begin{bmatrix} x_t^\top & u_t^\top \end{bmatrix} \right) \otimes \Sigma_w^{-1}\Bigg| \theta \right].
\end{align*}

\end{sidebar}

\paragraph{Regret Lower Bounds via Reduction to Bayesian Estimation}
To arrive at a local minimax lower bound (\ref{eq:genericlowerbound}) let us suppose for simplicity that $Q_T = P$, so that equation (\ref{eq:expectedregretrep}) becomes
\begin{equation}
\label{eq:regretlb1}
\begin{aligned}
     & \sup_{\theta \in B(\theta_\star,\e)} \E R_T^\pi(\theta)\\
      &= \sum_{t=0}^{T-1} \E_{\theta}^\pi \left[ (u_t - K(\theta)x_t)^\top (B^\top(\theta) P(\theta) B(\theta)+R)(u_t - K(\theta)x_t)\right]\\
    &\geq  \lambda_{\e} \sup_{\theta \in B(\theta_\star,\e)}    \sum_{t=0}^{T-1} \E_{\theta}^\pi  \|u_t - K(\theta)x_t\|_2^2.
\end{aligned}
\end{equation}
where $ \lambda_{\e}= \min_{\theta \in B(\theta_\star,\e)}\lambda_{\min}(B^\top(\theta) P(\theta) B(\theta)+R)\geq \lambda_{\min} (R)>0$. The next step is crucial: we relax the supremum in inequality (\ref{eq:regretlb1}) by a introducing a prior $\lambda$ over $\theta \in B(\theta_\star,\e)$. The exact choice of $\lambda$ is not particularly interesting and its influence on the final bound can be made to vanish. By weak duality we have for any such $\lambda$ that
\begin{equation}
\label{eq:regretlb2}
    \sup_{\theta \in B(\theta_\star,\e)} \E R_T^\pi(\theta)    \gtrsim   \sum_{t=0}^{T-1} \E_{\theta\sim\lambda} \E_{\theta}^\pi  \|u_t - K(\theta)x_t\|_2^2.
\end{equation}
The key insight is now that the quantity $\inf_{u_t}\E_{\theta\sim\lambda} \E_{\theta}^\pi  \|u_t - K(\theta)x_t\|_2^2$ is simply the MMSE for estimating the random variable $K(\theta) x_t$ where $\theta$ is drawn according to the prior distribution $\lambda$. Although it does require rather a few intermediate steps \citep[Theorem 4.1]{ziemann2022regret}, one can in principle lower bound  the right hand side of inequality (\ref{eq:regretlb2}) using estimation-theoretic lower bounds such as the Bayesian Cramér-Rao inequality \citep{gill1995applications}, namely \nameref{sidebar-vtineq}. The leading term in such lower bounds is the inverse of the Fisher Information:
\begin{align}
\label{eq:fisherinfoforldsintext}
     \I_p(\theta)= \E \left[\sum_{t=0}^{T-1}  \left( \begin{bmatrix} x_t \\ u_t \end{bmatrix}\begin{bmatrix} x_t^\top & u_t^\top \end{bmatrix} \right) \otimes \Sigma_w^{-1}\Bigg| \theta \right].
\end{align}
Heuristically, as $\e\to 0$, for two problem dependendent constants $c(\theta),c'(\theta)$, we have
\begin{align}
\label{eq:crudelb}
    \sup_{\theta \in B(\theta_\star,\e)} \E R_T^\pi(\theta) \geq  T \times c(\theta_\star) \lambda_{\min}  (\E_{\theta_\star}^\pi\I_p(\theta_\star)+c'(\theta_\star))^{-1}.
\end{align}
The reason the constant $ c(\theta_\star) $ is nonzero is a consequence of the derivative calcuation \eqref{eq:simchoprop}. This expression allows us to conclude that the jacobian terms discussed in \nameref{sidebar-vtineq} are invertible. Further, it is instructive to note that the expression inside the conditional expection in the expression (\ref{eq:fisherinfoforldsintext}) is proportional to the leading term in the estimation error (\ref{eq:LSEinAC}) related to recovery of the parameter $\theta=(A,B)$.

As we argued above following equation (\ref{eq:optpolsing}), the optimal policy $u_t = Kx_t$ renders the matrix (\ref{eq:fisherinfoforldsintext}) singular and so one needs to deviate from this policy to consistenly estimate the parameter $\theta=(A,B)$. In fact, it can be shown that the expected regret is an upper bound for the Fisher information (\ref{eq:fisherinfoforldsintext}): 
\begin{align}
\label{eq:crudeub}
     \lambda_{\min}(\E_{\theta}^\pi  \I_p(\theta) )\leq c''(\theta) \E R_T^\pi(\theta)
\end{align}
for a third problem dependent constant $c''(\theta)$, see \cite[Lemma 3.6]{ziemann2022regret}. This offers a slight change of perspective: the expected regret \eqref{eq:expectedregret} acts as a constraint on the set of possible experiment designs available to the learner. \rev{This idea has also been explored from the perspective of regret \emph{upper bounds} in \cite{colin2022regret}.}

Balancing the upper and lower bounds on the Fisher information in terms of the regret as in the heuristic inequalities \eqref{eq:crudelb}-\eqref{eq:crudeub}, yields that the optimal scaling must be $\sqrt{T}$. In particular, any policy attaining expected regret on the order of magnitude $O(\sqrt{T})$ generates a dataset where the smallest eigenvalue of the Fisher information is $O(\sqrt{T})$. Hence, identification of the parameter $\theta_\star=(A_\star,B_\star)$ can occur no faster than at the rate $O(1/\sqrt{T})$ for a regret-optimal policy, by which we can deduce that the optimal rate in fact is $\Omega(\sqrt{T})$. To obtain the correct dimensional dependence in the lower bound $\Omega(\sqrt{\dx \du^2 T})$, this argument needs to be slightly refined. Namely, we note that it in fact is not just the smallest eigenvalue of $\I_p(\theta)$ that is zero for laws of the form $u_t=Kx_t$ but in fact all the smallest $\dx\du$-many eigenvalues. To see this, note that the entire  linear manifold $\{(A,B) : A+BK_\star = A_\star +B_\star K_\star \}$, corresponds to parameters lacking persistency of excitation in closed-loop.

As mentioned above, the optimal dimensional scaling of regret for feedback systems has been settled by \cite{simchowitz2020naive}. However, there is currently a gap in our understanding of the best possible scaling of the regret in terms of key system-theoretic quantities. In particular, tight bounds for the scaling in terms the solution $P_\star$ to the steady state Riccati equation are unavailabe; the best known upper bound is due to \cite[Theorem 2]{simchowitz2020naive} and is of order $\sqrt{\opnorm{P_\star}^{11}}$, whereas the best known lower bound is of order $\sigma_{\min}(P_\star)$ \cite[Corollaries 4.2 and 4.3]{ziemann2022regret}. We note that ascertaining the exact optimal dependence of the regret on $P_\star$ and other system-theoretic quantities in LQR remains an open problem.

\paragraph{$P_\star$ can be exponential in the dimension}
We saw above that if one regards system-theoretic parameters as "dimension-less", the optimal dimensional-dependency for the state-feedback regret minimization scenario is polynomial in $\dx$ and $\du$. We will now see that these system-theoretic quantities can be rather significant.

To this end, consider the following system, which consists of two independent subsystems
\begin{equation}\label{CTRL_eq:REG_difficult_example_integrator}
A=\matr{{c|ccccc}0&0&0&&0&0\\\hline 0&1&1&&0&0\\& &&\ddots&\\0&0&0& &1&1\\0&0&0& &0&1},\,B=\matr{{c|c}1&0\\0&0\\\vdots\\0&1}.
\end{equation}
The first subsystem $(A_1,B_1)$ corresponding to the top and leftmost part of the arrays in equation (\ref{CTRL_eq:REG_difficult_example_integrator}) is just a simple memoryless system. The second subsystem $(A_2,B_2)$ is an integrator of order $\dx-1$. The system (\ref{CTRL_eq:REG_difficult_example_integrator}) is decoupled, but is very sensitive to miss-specification in their coupling due to the integrator component's potential for error amplification. Moreover, the solution $P_\star(A_2,B_2)$ is on the order $2^{\dx}$ \citep[Lemma 9]{Tsiamis2022Learning}. Using this one can construct a local minimax regret lower bound for the instance $(A,B)$ (system (\ref{CTRL_eq:REG_difficult_example_integrator})) with scaling
\begin{align*}
   \sup_{\theta \in B((A,B),\e)} \E R_T^\pi(\theta) = \Omega \left(  2^{\dx}\sqrt{T} \right).
\end{align*}
A more general statement  is given in \cite[Theorem 3]{Tsiamis2022Learning}. While the particular system \eqref{CTRL_eq:REG_difficult_example_integrator} has exponential complexity in the state dimension $\dx$ they establish a more general phenomenon: the controllability index $\kappa$---the number of steps it takes to reset a noise free system to the origin---can be used to characterize the local minimax regret and that this dependence is exponential. See also \autoref{table:adaptsum}. The discussion above leads to two conclusions:
\begin{enumerate}
    \item Learning to control can be hard; exponential complexity in the dimension can arise for examples as simple as integrators.
    \item To appreciate this hardness, we need to understand the role of control-theoretic quantities such as $P_\star$.
\end{enumerate}

\subsection{Partially Observed Systems}

While our current understanding of the state-feedback setting is relatively complete, less is known when the learner only has access to a measured output and not the actual system state. In the state-feedback setting, we know that the correct scaling with time is $\sqrt{T}$, that the dimensional dependence is $\sqrt{\dx \du^2}$ and that the key system-theoretic quantity is $P_\star$. By contrast, in the partially observed setting we currently only know that the correct scaling with the time horizon is $\sqrt{T}$. Determining the correct instance-specific scaling, and which quantities are key to this, is an open problem. Moreover, no current approach can handle the general LQG cost structure \eqref{eq:lqrcost} but instead apply to the criterion:
\begin{align*}
\tilde V_T^\pi(\theta) \triangleq  \E_\theta^\pi \left[\sum_{t=0}^{T-1} y_t^\top Q y_t + u_t^\top R u_t  \right].
\end{align*}
With these caveats in mind, we now sketch an elegant approach due to \cite{simchowitz2020improper} based on the classical Youla parametrization \citep{youla1976modern,zames1981feedback} leading to $\tilde O(\sqrt{T})$ regret for partially observed systems.

\subsection{Disturbance Feedback Control}
Unrolling the dynamics \eqref{eq:dynsysac}, it is straightforward to verify that 
\begin{align}
\label{eq:dynsysacunrolled}
    y_t = e_t + \sum_{s=0}^{t-1}C_\star A_\star^{t-s-1}w_s+\sum_{s=0}^{t-1}C_\star A_\star^{t-s-1}B_\star u_s
\end{align}
for some error signal $e_t$ decaying exponentially fast to $0$ for stable systems. The approach as sketched here  requires $\rho(A_\star)<1$ but can be extended to open-loop unstable systems \citep[Appendix C]{simchowitz2020improper}.

The representation \eqref{eq:dynsysacunrolled} suggests that there are two separate components to the input-output dynamics. The first component
\begin{align}
\label{eq:naty}
    y_t^{\mathsf{nat}} = e_t + \sum_{s=0}^{t-1}C_\star A_\star^{t-s-1}w_s
\end{align}
is referred to as "nature's $y$" and is a counterfactual object representing the evolution of the output in the absence of controller inputs. The second component is simply the discrete convolution of the inputs $u_{0:t-1}$ with the system Markov parameters $G^{0:t-1}_\star$ where $G_\star(s) = C_\star A_\star^{s}B_\star$. Hence $y_t = y_t^{\mathsf{nat}} + G^{0:t-1}_\star *u_{0:t-1} $. With these preliminaries established, for a sequence of matrices $\{M_s\}_{s=0}^{m-1}$ \cite{simchowitz2020improper} define disturbance response controllers (DRC) of order $m$ as controllers of the form
\begin{align}
\label{eq:dfc}
    u_t = \sum_{s=0}^{m-1} M_s y_{t-s}^{\mathsf{nat}}.
\end{align}
Notice that since $y_t^{\mathsf{nat}} = y_t - \sum_{s=0}^{t-1}C_\star A_\star^{t-s-1}B_\star u_s$, these are admissible causal controllers by construction --- had the dynamics $(A_\star,B_\star,C_\star)$ been known, we would have been able to execute controllers of the form \eqref{eq:dfc}. It can be shown that controllers of the form \eqref{eq:dfc} can approximate linear dynamic controllers such as the separation principle solution to LQG (Kalman filter with LQR controller). 
\subsection{Regret Bounds for Partially Observed Systems}
The following algorithm combines the convex Youla-like parametrization \eqref{eq:dfc} with modern Online Convex Optimization \citep{anava2015online}. In particular, \citet{simchowitz2020improper} propose an algorithm  in which they:
\begin{enumerate}
    \item inject exploratory noise for a period of length proportional to $\sqrt{T}$;
    \item use this dataset to estimate the Markov parameters $M$;
    \item for the remainder of the horizon compute estimates of nature's $y$ \eqref{eq:naty} using the estimated Markov parameters; and
    \item use the estimated nature's $y$ to run online (projected) gradient descent on the parameters $M_s$ of the disturbance feedback controller.
\end{enumerate}
\citet{simchowitz2020improper} show that for a properly tuned order $m$ of DRC the approach outlined above yields $\tilde O (\sqrt{T})$ regret. While in this setting there is no general lower bound to date, \cite{ziemann2022regret} have shown that $\Omega(\sqrt{T})$ regret is unavoidable in the worst case by considering instances with large input dimension. 

\paragraph{Logarithmic Regret?}
It is also interesting to note that for an alternative notion of regret, in which the learner competes with the best \emph{persistently exciting} policy instead of the optimal policy, \cite{lale2020logarithmic} has shown that logarithmic regret is possible in the partially observed setting. We note however that the optimal LQG policy might not necessarily be persistently exciting. Indeed, known lower bounds show that it is not persistently exciting in i) the state-feedback setting (cf. \eqref{eq:optpolsing}); and ii) the partially observed setting for certain large input dimension systems. Thus, it is an open problem to characterize the relation between the regret definition~\eqref{eq:regret} and the one defined in~\cite{lale2020logarithmic}. 

We note in passing that a related situation arises in the state-feedback setting if the learner is given access to the precise value of $B_\star$. In this case, it suffices to identify the matrix $A_\star$, which is identifiable in closed-loop given knowledge of $B_\star$. \citet{cassel2020logarithmic} show that this observation leads to logarithmic regret---against the optimal controller---if $B_\star$ is known a priori. 

A related problem where logarithmic regret is possible is that of adaptive Kalman filtering or online prediction~\cite{kozdoba2019line,tsiamis2020online,ghai2020no,rashidinejad2020slip}. The objective is to predict future observations $y_k$ online based on the past $y_{k-1},u_{k-1},\dots,y_0,u_0$. Since the only goal is prediction, the cost of control does not enter the objective. Interestingly, for this problem it is possible to attain logarithmic regret~\cite{tsiamis2020online,ghai2020no,rashidinejad2020slip}. 
Hence, we can learn the Kalman filter online with a smaller regret than that achievable in online LQR control. In light of our discussion, this is hopefully no longer surprising. In the LQR problem, we need to inject additional exploratory signals into the system, which also affects the cost of control. In the prediction problem, exploration is ``free'' as the cost of control does not affect prediction performance. In fact, we can predict even without persistence of excitation~\cite{ghai2020no}; informally, if the covariates lie on a certain subspace, so will  their future versions.

\begin{open_prob}
\label{op:exactratelqrregret}
Provide matching upper and lower bounds on either the regret \eqref{eq:regret} or the expected regret \eqref{eq:expectedregret}. In the partially observed setting, we currently do not even know the correct dimensional-dependence (or what the correct notion of dimension is---although it is to be suspected that this is related to the order of the system and the input and output dimensions $\du$ and $\dy$).

To resolve this problem it is required to find a function $f$ such that for a universal constant $c_1>0$ independent of all problem parameters, we have:
\begin{align*}
     \Regret_T^{\pi}(A_\star,B_\star) &\leq c_1 f(A_\star,B_\star,C_\star,Q,R,\Sigma_W,\Sigma_V,T) 
\end{align*}
for some some specific algorithm $\pi$ and for $T$ sufficiently large with high probability (or in expectation). A resolution will also provide a matching lower bound, which for some $\e=o_T(1)$ and some constant $c_2>0$ only depending on $\e$ establishes that:
\begin{align*}     
     \sup_{A,B\in B((A_\star,B_\star),\e)}\Regret_T^\pi(A,B) &\geq c_2 f(A_\star,B_\star,C_\star,Q,R,\Sigma_W,\Sigma_V,T) 
\end{align*}
for all algorithms $\pi$ and for $T$ sufficiently large with at least constant probability (or in expectation). A partial resolution only applying to state-feedback systems, thus determining the correct dependence on system-theoretic quantities is also of interest.
\end{open_prob}

\section{Summary and Discussion}
We have provided a tutorial survey of recent advances in statistical learning for control. One of the key takeaway messages is that we now have a relatively complete picture of the learning problem in fully observed linear dynamical systems, both in terms of  system identification, as summarized in \autoref{ID_TAB:Sys_ID_Fully}, and in terms regret minimization as summarized in \autoref{table:adaptsum}. We have also provided an overview and listed a number of open problems in particular with respect to partially observed extensions of the above-mentioned results. Indeed, as exciting as the developments over the past few years in this field have been, there is still much work to be done. With this mind, we now outline some future directions we believe are important for the field to consider as next steps. 

\section{future directions}
\label{sec:futdir}
\subsection{Control Oriented Identification}
In~\nameref{sec:finsampsysid}, we studied methods of obtaining high probability bounds on the parameter estimation error of the form
\[
\opnorm{\hat{A}_T-\As}\le \varepsilon,
\]
where $\hat{A}_T$ is the output of the least squares algorithm~\eqref{ID_eq:ls_algorithm}. Similar bounds can be obtained for the other state parameters as well. As we discussed
in~\nameref{sec:ellipsoids}, the operator norm picks up the worst-case direction which is the most difficult to identify. In fact, as shown in~\cite{tsiamis2021linear}, the sample complexity of identifying the worst-case direction can grow very large for certain systems. 
However, a question that arises is whether this worst-case direction affects control. ``\emph{Does the bottleneck of identification, i.e., the worst direction, affect control design? Do we always need to identify everything?}"

\rev{Consider for example the following system
\[
A_1=\begin{bmatrix} 0&\alpha&0\\0&0&\beta\\0&0&0\end{bmatrix},\,B=\begin{bmatrix} 0\\0\\1\end{bmatrix},\,\Sigmaw=\begin{bmatrix} 1\\0\\0\end{bmatrix}\begin{bmatrix} 1\\0\\0\end{bmatrix}^\top,
\]
where only $\alpha$ and $\beta$ are unknown. Let the control objective be stabilization by state feedback, i.e. finding a feedback gain $K$ such that the closed-loop system $A+BK$ is asymptotically stable.
 The only way to excite $x_{k,2}$ is via $x_{k,3}$; the coupling coefficient $\beta$ determines the degree of excitation. Note that as the coupling $\beta$ goes to zero, the excitation of $x_{k,2}$ becomes smaller and smaller. As a result, if $\beta$ is very small it is very difficult to identify the parameter $\alpha$ and the complexity of system identification increases with $\beta^{-1}$. However, it is trivial to stabilize the system, even without knowledge of $\alpha$ e.g. with $K=0$. In this particular example, the worst direction of identification error is not relevant for stabilization. Hence, the complexity of stabilization should be independent of $\beta^{-1}$.}

\rev{On the other hand, consider system
\[
A_2=\begin{bmatrix} 1&\alpha&0\\0&0&\beta\\0&0&0\end{bmatrix},\,B=\begin{bmatrix} 0\\0\\1\end{bmatrix},\,\Sigmaw=\begin{bmatrix} 1\\0\\0\end{bmatrix}\begin{bmatrix} 1\\0\\0\end{bmatrix}^\top,
\]
where now the first state has marginally stable dynamics. Unfortunately, for this pathological example, it is in fact necessary to identify $\alpha$ in order to stabilize the system (this example is adapted from~\cite{Tsiamis2022Learning}), suffering from complexity which scales with $\beta^{-1}$.} In particular, we cannot stabilize the system unless we identify the sign of $\alpha$, showing that for some systems, the worst direction of the identification error matters. The example above shows a system for which stabilization depends on an \emph{identification bottleneck}. 
However, it seems that the constructed systems are artificial or pathological. It is an open problem to characterize the conditions under which we can avoid such corner cases.

Similar questions have been previously studied in the context of control-oriented identification or identification for control~\cite{gevers2005identification}. In many situations of practical interest we only need to identify the part of the model that matters for a specific closed-loop objective. In this case, it is reasonable to 
tune the identification towards the objective for which the model is to be used, i.e., to ensure that the model error is ``orthogonal'' to the control objective. 
This is particularly important in the case of agnostic learning, i.e., when there is no ``true model" and the model class can only approximate the system, which is typically the case in practice.

\subsection{Learning with Structure and Regularization}
In many practical situations, certain structural properties of the system to be identified and controlled are known a priori. For instance, when trying to learn a networked system, the engineer might have prior knowledge that interconnections between states are relatively sparse. Other examples of relevant structural priors include low order, as captured by the rank of a system Hankel matrix, or physical properties such as passivity and dissipativity.

\paragraph{Sparsity}
 In the case of a linear dynamical system, sparsity amounts to the matrix $A_\star$ in the dynamics $x_{t+1}=A_\star x_t +w_t$ having many zero entries, i.e., $A_\star$ will be sparse and have only $s\ll \dx^2$ nonzero entries. Many modern networked systems have the property that they are large scale but not maximally connected, leading to a high-dimensional state vector with sparse $A_\star$. There are many other examples that fall into this category, including snake-like robots, which can be modeled by an integrator-like structure:
 \begin{align*}
     A_{\mathsf{snake}}= \begin{bmatrix}
     a_{11}& a_{12} & 0 & 0 &\dots & 0\\
     0 & a_{22} & a_{23} &0 &\dots &0 \\
     \vdots & \ddots &\ddots &\ddots & \ddots &\vdots
     \end{bmatrix}.
 \end{align*}
The matrix $A_{\mathsf{snake}}$ has only $s=2\dx\ll \dx^2$ many nonzero entries and so one is justified to hope for a polynomial speed-up in the sample complexity of system identification as compared to the standard minimax rate achieved by the least squares estimator.

 In such high-dimensional situations, running linear regression, which suffers a minimax rate of convergence proportional to the $\dx^2$ in Frobenius norm (proportional to $\dx$ in operator norm), is not sample efficient or might not even be tractable. To alleviate this issue,  \citet{fattahi2019learning} analyze the LASSO estimator as applied to system identification. Recall that the $\ell^1$-norm of a vector $v = (v_1,\dots,v_\mathsf{d}) \in \mathbb{R}^{\mathsf{d}}$ takes the form $\|v\|_{\ell^1} =\sum_{i=1}^\mathsf{d} |v_i| $. The LASSO penalizes the least squares solution by this norm using a fixed regularization parameter $\lambda>0$, and takes the form:
 \begin{align}
 \label{eq:thelasso}
     \widehat A \in \argmin_{A\in \mathbb{R}^{\dx}} \left\{ \frac{1}{T}\sum_{t=0}^{T-1} \|x_{t+1}+Ax_t\|_2^2 + \lambda \|\VEC(A)\|_{\ell^1} \right\}.
 \end{align}
 It is by now well known that $\ell^1$-regularization promotes sparse least squares solutions \cite{bickel2009simultaneous, negahban2009unified}. Indeed, the authors of \cite{fattahi2019learning} show that the LASSO also avoids polynomial dependence on the state dimension for linear dynamical systems. Unfortunately however, the rate in \cite{fattahi2019learning} degrades with the stability of the system---precisely that which we sought to avoid in our discussion of \nameref{sec:finsampsysid} by leveraging \nameref{sidebar-small-ball}. Moreover, by instantiating recent results in \cite{ziemann2022learning} it can be shown that the minimax rate (in Frobenius norm) over the class of $s$-sparse linear dynamical systems is no more than $\tilde O\left(\sqrt{\frac{s \sigma_w^2}{\lambda_{\min}(\Gamma_T)} }\right)$ where $\Gamma_T$ is as in \eqref{eq:gramdef} (with $B_\star =0)$. Unfortunately, instantiating  \cite{ziemann2022learning} does not yield an effective algorithm and reduces to running ${\dx^2 \choose s} =O( \dx \exp (2s))$ separate regressions, each one over an $s$-dimensional sub-manifold. This quickly becomes intractable even for rather moderate cases of the degree of sparsity  $s$.
 
 \begin{open_prob}
 
 Studying the tension between dependence on mixing time (stability) and computational intractability is an exciting direction for future work. Can we refine existing analysis of the LASSO (or provide some other polynomial algorithm) to match minimax rates, or is there a fundamental computational barrier introduced by sparsity? Resolving this issue may well require the development of new tools since existing analyses of the LASSO in the i.i.d. setting invariably  depend on the condition number of the covariates matrix \citep{lecue2018regularization, wainwright2019high}, which for a linear dynamic system is proportional to the mixing time (degree of stability), leading to sub-optimal rates.
 
 \end{open_prob}
 \paragraph{Low Order Models}
 
Sparsity as discussed above is also relevant when estimating input-output models of unknown order. For example, consider the following model:
\begin{align}
\label{eq:unknownorderio}
    y_{t+1}= \sum_{j=0}^t A_j y_{t-j}+ \sum_{j=0}^t B_j u_{t-j}+w_t, && y_j = 0 \textnormal{ for } j \leq 0.
\end{align}
In this scenario, there is no nontrivial upper bound on the lag order available to the engineer, and it may be as large as the entire horizon $T$. Converting the process \eqref{eq:unknownorderio} into state space form and running least squares is not tractable: recall that the minimax rate of convergence depends on the ratio of the number of unknown parameters and the number of samples (in this case, given by the horizon $T$). Without further assumption this ratio is constant in the worst-case for model \eqref{eq:unknownorderio}. However, if there is hope that the true model is of low order so that many of the $\{A_j,B_j\}$ are zero, a variation of the LASSO \eqref{eq:thelasso} may also be appropriate for model selection in this scenario.

\paragraph{Low Rank Models}
A more sophisticated notion of model order than discussed in the preceding paragraph is that of Hankel matrix rank (McMillan degree). Let $h_\star = \begin{bmatrix}  C_\star B_\star & C_\star A_\star B_\star & C_\star A^2_\star B_\star & \dots \end{bmatrix}$ denote the impulse response (matrix) associated to the tuple $(A_\star,B_\star,C_\star)$ and notice that model \eqref{ID_eq:system} can be written as
\begin{align*}
    y_t = h * u_{0:t-1}+\eta_t
\end{align*}
where $*$ denotes discrete convolution, and $\{\eta_t\}$ is some (not necessarily i.i.d.) noise sequence. Denote by $\mathcal{H}$ the Hankel (linear) operator, mapping impulse responses to Hankel matrices.  The nuclear norm of a matrix $M \in \mathbb{R}^{\mathsf{d}\times \mathsf{d}}$ is $\|M\|_* = \sum_{i=1}^\mathsf{d} \sigma_{i}(M)$. This norm plays a similar role to the $\ell^1$-norm but promotes low rank solutions rather than sparse solutions \cite{negahban2009unified}. Since the rank of the Hankel matrix $\mathcal{H}(h_\star)$ coincides with the McMillan degree of the system \eqref{ID_eq:system}, it is natural to consider the following nuclear norm regularized problem (see e.g. \cite{sun2022system}):
\begin{align}
 \label{eq:thenuc}
     \widehat h \in \argmin_{h} \left\{ \frac{1}{T}\sum_{t=0}^{T-1} \|y_{t}+h*u_{t-1:0}\|_2^2 + \lambda \|\mathcal{H}(h)\|_{*} \right\}.
 \end{align}
 As of the writing of this article, no finite sample analysis exists for the nuclear norm regularized estimator \eqref{eq:thenuc}.

\subsection{Learning for Nonlinear Identification and Control}

While the vast majority of the literature on statistical learning for identification and control has been on linear systems, most real systems are not. Learning in linear dynamical systems escapes many nonlinear phenomena and does not capture one of the most fundamental issues in modern machine learning, distribution shift. For linear models, parameter recovery is always possible as long as the average covariance matrix of the covariates is sufficiently non-degenerate (invertible) and the rate of parameter recovery is (asymptotically) completely described by the second order statistics of the process under investigation. Put differently, all equilibrium points of a linear system are (dynamically) equivalent. This stands in stark contrast to more general nonlinear systems in which, in the worst case, learning the behavior around one equilibrium point gives no information about the behavior of the system in other regions of the state-space.  

Moreover, recent advances in learning and estimation for nonlinear dynamics bypass these issues of distribution shift by either considering models which behave almost linearly \cite{sattar2020non, mania2022active, foster2020learning,sattar2021identification,  kowshik2021near} or by sidestepping the issue entirely and only considering a prediction error associated to the invariant measure of the system \cite{ziemann2022single, ziemann2022learning}. For statistical learning to be truly informative for downstream control applications a more integrated understanding of learnability, nonlinear dynamic phenomena, and control-theoretic notions such as incremental stability or contraction are needed \citep{tu2022sample, pfrommer2022tasil, tsukamoto2021contraction}.

\paragraph{Realizability and Approximation}
Existing work on learning in dynamical systems make strong realizability assumptions. For instance, it is often assumed that the true model is generated by a linear dynamical system of the form \eqref{ID_eq:system} driven by i.i.d. mean zero (or martingale difference) noise. Even if one considers more complicated nonlinear models, such additive mean zero noise models completely sidestep bias or misspecification challenges.

This is significant since ignoring this issue might mean that existing analyses are overly optimistic. Indeed, \cite{nagaraj2020least} shows that in the worst case, misspefication in a simple linear regression model leads to a deflated sample complexity by a factor linear in the mixing time of the covariates process. This stands in stark contrast to the results in \cite{simchowitz2018learning} in which linear regression over a well-specified model class is analyzed completely without reference to mixing. While the fundamental limits in \cite{nagaraj2020least} may seem discouraging at first, they are worst case, and may be avoidable by introducing further regularity assumptions. As a first step, one could analyze the sample complexity of recovering the best linear approximation to an almost linear autoregression, e.g., adding a small nonlinearity, or considering a generalized linear model with nearly isometric link function.

\paragraph{Structured Nonlinear Identification}
A host of new opportunities present themselves in structural nonlinear identification as compared to the linear setting. While, sparse and low-rank structure are certainly of interest and applicable to learning in nonlinear dynamical systems, there are other exciting, and arguably more fundamentally system-theoretic, alternatives. For instance, one might ask how properties such as passivity or dissipativity affect the minimax rate of estimation and whether there are efficient algorithms that might take advantage of this. More concretely, one might be interested in the  $1$-dimensional autoregression $x_{t+1} = f_\star(x_t)+w_t$ and seek to identify $f_\star$ under the physically motivated hypothesis that $f_\star$ is the negative gradient of an unknown convex potential. 

Taking advantage of structure may also be inherently more important in nonlinear identification since otherwise the curse of dimensionality is quick to present itself. For instance, in the model
\begin{align*}
    y_t = f_\star(x_t) +w_t
\end{align*}
running regression over the hypothesis class $\mathscr{F} = \{ f : \mathbb{R}^{\dx} \to [0,1] \subset \mathbb{R} \textnormal{ and $f$ is $k$-smooth}\}$ incurs a minimax rate which degrades \emph{exponentially} with large $\dx$.

\section{ACKNOWLEDGMENT}

The work of N. Matni is supported in part by NSF award CPS-2038873, NSF CAREER award ECCS-2045834,
and a Google Research Scholar award. The authors are grateful to three anonymous reviewers for excellent feedback.

\section{Author Information}
\begin{IEEEbiography}{Anastasios Tsiamis}{\,}(Member, IEEE) received the
Diploma degree in electrical and
computer engineering from the
National Technical University of
Athens, Greece, in 2014, and  a Ph.D. in Electrical and Systems
Engineering at the University of
Pennsylvania, in 2022.  He is currently a postdoctoral scholar with Department of Information Technology and Electrical Engineering at ETH Zürich. His research interests
include statistical learning for control, risk-aware control and optimization, and networked control systems. Anastasios Tsiamis was
a finalist for the International Federation of Automatic Control (IFAC) Young Author Prize in IFAC 2017 World Congress and a finalist for the Best Student Paper Award in American Control Conference (ACC) 2019.
\end{IEEEbiography}

\begin{IEEEbiography}{Ingvar Ziemann}{\,}(Student Member, IEEE) received his PhD in November 2022 from the Division of Decision and Control Systems at The Royal Institute of Technology (KTH) under the supervision of Henrik Sandberg. His research is centered on using statistical and information theoretic tools to study learning-enabled control methods, with a current interest in studying how learning algorithms generalize in the context of dynamical systems. Prior to starting his Ph.D., he obtained two sets of Master's and Bachelor's degrees in Mathematics (SU/KTH) and in Economics and Finance (SSE). Ingvar is the recipient of a Swedish Research Council International Postdoc Grant, the IEEE CDC 2022 Best Student Paper Award, and the 2017 Stockholm Mathematics Center Excellent Master Thesis Award.
\end{IEEEbiography}

\begin{IEEEbiography}{Nikolai Matni}{\,}(Member, IEEE) is an Assistant Professor in the Department of Electrical and Systems Engineering at the University of Pennsylvania, where he is also a member of the Department of Computer and Information Sciences (by courtesy), the GRASP Lab, the PRECISE Center, and the Applied Mathematics and Computational Science graduate group. Prior to joining Penn, Nikolai was a postdoctoral scholar in EECS at UC Berkeley. He has also held a position as a postdoctoral scholar in the Computing and Mathematical Sciences at Caltech. He received his Ph.D. in Control and Dynamical Systems from Caltech in June 2016. He also holds B.A.Sc. and M.A.Sc. in Electrical Engineering from the University of British Columbia, Vancouver, Canada. His research interests broadly encompass the use of learning, optimization, and control in the design and analysis of safety-critical data-driven autonomous systems. Nikolai is a recipient of the NSF CAREER Award (2021), a Google Research Scholar Award (2021), the 2021 George S. Axelby award, the IEEE ACC 2017 Best Student Paper Award (as co-advisor), and the IEEE CDC 2013 Best Student Paper Award. 
\end{IEEEbiography}

\begin{IEEEbiography}{George J. Pappas}{\,} (Fellow, IEEE) received the Ph.D. degree in electrical engineering and computer sciences from the University of California, Berkeley, CA, USA, in 1998. He is currently the Joseph Moore Professor and Chair of the Department of Electrical and Systems Engineering, University of Pennsylvania, Philadelphia, PA, USA. He also holds a secondary appointment with the Department of Computer and Information Sciences and the Department of Mechanical Engineering and Applied Mechanics. He is a member of the GRASP Lab and the PRECISE Center. He was previously the Deputy
Dean for Research with the School of Engineering and Applied Science. His research interests include control theory and, in particular, hybrid systems, embedded systems, cyber-physical systems, and hierarchical and distributed control systems, with applications to unmanned aerial vehicles, distributed robotics, green buildings, and biomolecular networks. Dr. Pappas was a recipient of various awards, such as the Antonio Ruberti
Young Researcher Prize, the George S. Axelby Award, the Hugo Schuck Best Paper Award, the George H. Heilmeier Award, the National Science Foundation PECASE award, and numerous best student papers awards. 
\end{IEEEbiography}

\bibliographystyle{unsrtnat}
\bibliography{main.bib}

\endarticle
\end{document}